\documentclass[usenatbib,usegraphicx]{mn2e}
\usepackage{times}
\usepackage{subfigure}
\usepackage{amssymb}
\newcommand{\Rsun}{\hbox{R$_{\odot}$\,}}
\newcommand{\Msun}{\hbox{M$_{\odot}$\,}}
\newcommand{\Rj}{\hbox{R$_{Jup}$\,}}
\newcommand{\dchs}{\hbox{$\Delta\chi^2$}}
\newcommand{\SW}{\hbox{SuperWASP}}
\newcommand{\ntr}{58} 
\newcommand{\ntru}{55} 

\title[Extrasolar Planet Candidates from SuperWASP-North]{SuperWASP-North Extrasolar Planet 
Candidates. Candidates from Fields 17hr $<$ RA $<$ 18hr}
\author[T.A.~Lister et al.]{T.A.~Lister$^{1,2,12}$\thanks{Current address: Las
Cumbres Observatory. E-mail: tlister@lcogt.net},
R.G.~West$^{3}$,
D.M.~Wilson$^{2}$,
A. Collier Cameron$^{1}$,
W.I. Clarkson$^{4,8}$,
\newauthor
R.A. Street$^{5,13}$,
B.~Enoch$^{4}$,
N.R. Parley$^{4}$,
D.J. Christian$^{5}$,
S.R.~Kane$^{1,9}$,
A. Evans$^{2}$,
\newauthor
A. Fitzsimmons$^{5}$,
C.A. Haswell$^{4}$,
C. Hellier$^{2}$,
S.T. Hodgkin$^{6}$, 
Keith Horne$^{1}$,
\newauthor
J. Irwin$^{6}$,
F.P. Keenan$^{5}$,
A.J. Norton$^{4}$,
J. Osborne$^{3}$,
D.L. Pollacco$^{5}$,
R. Ryans$^{5}$,
\newauthor
I. Skillen$^{7}$,
P.J. Wheatley$^{10}$
and J. R. Barnes$^{1,11}$ \\
$^{1}$SUPA, School of Physics \& Astronomy, University of St. Andrews, North Haugh, 
St. Andrews, Fife, KY16 9SS, UK,  \\
$^{2}$Astrophysics Group, School of Chemistry \& Physics, Keele University, 
Staffordshire, ST5 5BG, UK,  \\
$^{3}$Department of Physics \& Astronomy, University of Leicester, Leicester, 
LE1 7RH, UK,\\
$^{4}$Department of Physics \& Astronomy, The Open University, Milton Keynes, 
MK7 6AA, UK,  \\
$^{5}$Astrophysics Research Centre, Main Physics Building, School of Mathematics
and Physics, Queen's University Belfast, BT7 1NN, UK,\\ 
$^{6}$Institute of Astronomy, University of Cambridge, Madingley Road, 
Cambridge, CB3 0HA, UK, \\ 
$^{7}$Isaac Newton Group of Telescopes, Apartado de correos 321,
E-38700 Santa Cruz de la Palma, Tenerife, Spain,\\ 
$^{8}$Space Telescope Science Institute (STScI), 3700 San Martin Drive,
Baltimore, MD~21218, USA,\\
$^{9}$University of Florida, PO~Box~112005, 211 Bryant Space Science Center,
Gainesville, FL, USA.\\
$^{10}$Department of Physics, University of Warwick, Coventry CV4 7AL, UK.\\
$^{11}$Centre for Astrophysics Research, Science \& Technology Research
Institute, University of Hertfordshire, Hatfield, AL10 9AB\\
$^{12}$Las Cumbres Observatory, 6740B Cortona Drive, CA 93117, USA.\\
$^{13}$Dept. of Physics, Broida Hall, University of California, Santa
Barbara, CA 93106-9530, USA.\\} 

\date{Accepted 2007 ?? ??. Received 2006 ??; in original form 2006 
 ??}

\pagerange{\pageref{firstpage}--\pageref{lastpage}} \pubyear{2006}

\begin{document}
\maketitle
\label{firstpage}

\begin{abstract}
We have performed photometric observations of nearly 7 million stars with $8 < V
< 15$ with the \SW-North instrument from La Palma between 2004 May--September.
Fields in the RA range 17--18\,hr, yielding over 185,000 stars with sufficient
quality data, have been searched for transits using a modified box least-squares
(BLS) algorithm. We find a total of \ntr\  initial transiting candidates which
have high S/N in the BLS, show multiple transit-like dips and have passed visual
inspection. Analysis of the blending and inferred planetary radii for these
candidates leaves a total of 7 transiting planet candidates which pass all the
tests plus 4 which pass the majority. We discuss the derived parameters for
these candidates and their properties and comment on the implications for future
transit searches.
\end{abstract}


\section{Introduction}

Since the discovery of the first extrasolar planets around pulsars
\citep{wolszczan92} and Sun-like stars \citep{mayor95} and the subsequent
discovery of over 200 other planets\footnote{http://exoplanet.eu/catalog.php}
in the following decade, many questions about their formation, evolution and
distribution have been raised. In particular the discovery of the ``hot
Jupiters'' with orbital periods less than 5 days has produced a large
transformation in the theory of planetary formation and migration to explain
how objects can be formed far out in the protoplanetary nebula and brought
through the disc and stopped at the very small orbital radii observed.

Although the vast majority of extrasolar planets have been discovered using the
radial velocity technique, it is the small subset that transit their parent star
that have the greatest potential as these are the only ones for which masses and
radii can be determined without the $\sin i$ ambiguity. In addition,
observations of transits have been used to examine the atmosphere
\citep{charbonneau02} and evaporating exosphere (\citealt{vidal03},
\citealt{vidal04}) of exoplanets, to search for moons and rings \citep{brown01},
while long-term observations of transits have the potential to reveal other
planets down to Earth masses \citep{agol05}.

\begin{table*}
\centering
\caption{Journal of spectroscopic observations of transit candidates at the CFHT.}
\protect\label{tab:sjournalcfht}
\vspace{5mm}
\begin{tabular}{lccccl}
\hline
Object (1SWASP+)   & UT start & Exp time (s) & Raw S/N & Deconvolved S/N & Comments \\
\hline
\multicolumn{6}{c}{2005 September 23} \\
\hline
J174645.84+333411.9 & 06:12 & 600 & $\sim 30$ & $\sim 820$ & Seeing $\sim1.1$\arcsec \\
\hline
\multicolumn{6}{c}{2005 September 24} \\
\hline
J174645.84+333411.9 & 05:45 & 400 & $\sim 30$ & $\sim 795$  & Seeing $\sim0.8$\arcsec \\
J173403.61+280145.1 & 06:51 & 500 & $\sim 26$ & $\sim 720$  & Seeing $\sim0.5$\arcsec \\
J172826.46+471208.4 & 07:03 & 300 & $\sim 16$ & $\sim 430$  & Seeing $\sim0.5$\arcsec \\
\hline
\end{tabular}
\end{table*}

Since the first discovery of the transiting extrasolar planet (HD~209458b;
\citealt{charbonneau00}), many other searches have been instigated with the aim
of discovering transiting planets (see \citealt{horne03} for a review). The low
cost of the equipment used to detect HD~209458b and its easily detected, large
inflated radius, which still challenges exoplanetary atmospheric theories (e.g.
\citealt{burrows03}, \citealt{laughlin05}), have led to an underestimation of
the difficulties of the data reduction needed to reach the required precision
over very wide fields (see discussion by \citealt{bakos04hat}).

The first extrasolar planets to be discovered by the transit method were found
by the OGLE project in 2002 (\citealt{udalski02}; \citealt{udalski02extras};
\citealt{udalski02carina}) and five of the systems have been confirmed
spectroscopically. However this task is made more difficult by the faintness
($V\simeq15$--18) of these stars, necessitating large amounts of time on very
large telescopes. Detailed follow-up studies such as atmospheric spectroscopy
and Spitzer secondary eclipse detections (\citealt{charbonneau05spitzer},
\citealt{deming05spitzer}, \citealt{deming06spitzer}) require brighter targets.
The detection of TReS-1 \citep{alonso04} was the first of an extrasolar planet
around a bright star from a ``shallow and wide'' survey. 

The goal of the WASP Project and the \SW\  instruments is to provide a large
number of bright ($9 < V < 13$) extrasolar planet candidates to allow meaningful
statistical studies to be carried out and allow follow-up to be undertaken with
telescopes of moderate aperture. This paper is part of a series (following
\citealt{christian06}) describing results of a search for transiting extrasolar
planets from the first season of operations in 2004.

Section~\ref{sec:obsinst} describes the instrumentation and the observing
strategy and the limited spectroscopic follow-up. The data reduction, pipelining
and archive extraction stages are described in Section~\ref{sec:data}. The
transit search, candidate filtering and selection procedure is described in
Section~\ref{sec:select}  and results of the search are discussed in
Section~\ref{sec:results}. Finally we summarize our findings in
Section~\ref{sec:discuss} and give conclusions in Section~\ref{sec:conclusion}.

\section{Observations and instrumentation}
\protect\label{sec:obsinst}

\subsection{Photometry}
\protect\label{sec:obsphot}
The photometric data were obtained with the SuperWASP-North instrument at the
Observatorio del Roque de los Muchachos, La Palma, Canary Islands, during 2004
May--September. At this time the instrument consisted of five cameras guided by
an equatorial fork mount made by Optical Mechanics Inc. Each camera was made up
of a Canon 200mm, f/1.8 telephoto lens coupled to a Andor
Technologies\footnote{http://www.andor-tech.com} 2048$\times$2048 pixel CCD
camera which uses a Marconi (now e$^2$v) thinned, back-illuminated CCD with
13.5$\mu$m pixels with Peltier thermoelectric cooling. The resulting plate
scale is 13.7\,arcsec/pix, and each camera has a field of view of
7.8$^{\circ}\times$7.8$^{\circ}$.  The equipment is described in greater detail
in \citet{pollacco06}.

The observational strategy was designed to primarily target a band of fields at
$\rmn{Dec}=+28^\circ$ (corresponding approximately to the latitude of La Palma)
at 1 hour increments in Right Ascension. The individual cameras were offset from
this position by approx. $\pm3.5$\,deg in RA and Dec. The fields were chosen to
avoid the densest part of the Galactic plane with no fields at RA=19--20\,h.
This prevented source confusion and blending with our large pixel scale which
has been shown \citep{brown03} to be a significant source of false positives in
wide-field transit surveys. Fields within $30^\circ$ of the Moon were not
observed and 8--14 fields were observed each night with 30\,s exposures at a
cadence of $\sim8$ mins.

In all a total of 165 fields were observed with a variable number of
observations per field on 135 nights giving a total of $\sim12.9$ billion
photometric data points being obtained on $\sim6.7$ million unique objects. This
number does not include the ``orphans''; objects detected in the individual CCD
frames but which are not present in the photometric catalogue (see
Section~\ref{sec:pipeline}) which have been excluded from the analysis.

\subsection{Spectroscopy}
\protect\label{sec:obsspec}
Follow-up spectroscopic observations of a small number of transit candidates
were obtained by one of us (JRB) as part of another observing program using the
3.6m Canada France Hawaii Telescope (CFHT) and ESPaDOnS echelle spectrograph
(\citealt{donati97zdi}, \citealt{donati03espadons}) on Mauna Kea, Hawaii on the
nights of 2005 September 23--24. The instrument was configured in
spectropolarimetric mode using the 79 gr/mm grating and the
$2\rmn{k}\times4.5\rmn{k}$ pixel EEV1 CCD detector, giving  $R\sim63000$ and a
wavelength coverage over 40 orders of 370--1050\,nm. Observations were obtained
in Stokes I and exposure times ranged from 300 to 600\,s depending on the
brightness of the target. A journal of the spectroscopic observations is shown
in Table~\ref{tab:sjournalcfht}.

The data were reduced automatically at the telescope using
\textit{Libre-ESpRIT}\footnote{\hbox{http://www.cfht.hawaii.edu/Instruments/Spectroscopy/Espadons/Espadons\_esprit.html}}
to perform bias subtraction, flat-fielding, wavelength calibration and order
extraction of the polarization information. The reduction process also makes use
of the telluric water lines within the echellogram to align the velocity scale
to within a few tens of m/s of the heliocentric reference frame.

The extracted spectra were then analysed using the technique of Least Squares
Deconvolution (\citealt{donati97doppler}, \citealt{donati97zdi}, Appendix C of
\citealt{cameron02upsand}) to boost the S/N of the spectra. Least Squares
Deconvolution (LSD) makes use of the large number of images of photospheric
lines recorded in the several hundred nanometres of wavelength range covered by
an echelle spectrograph to increase the S/N by a factor $\sim \sqrt{\rmn{no.\
of\ line\ images}}$. We used a solar G2 line list in the deconvolution process
and obtained 4688 images of 3507 lines for the three targets, giving an
increase in S/N of $\sim 27$.

\section{Data analysis}
\protect\label{sec:data}

\subsection{Data reduction \& pipelining}
\protect\label{sec:pipeline}

The photometric data were reduced using the WASP0/SuperWASP automated pipeline
(\citealt{pollacco06}). Frames are initially classified through a series of
statistical tests into bias, dark flat, object and defective frames. Tests
specific to each calibration frame type such as the readout noise (for bias
frames), dark current (for darks) and the number of saturated pixels and the
illumination gradient (for flat fields) are carried out to determine their
usability.

Bias and dark frames are then optimally combined using \textsc{ccdpack}
\citep{ccdpack} into nightly master calibration files. The automated sequences of
flat fields obtained at dusk and dawn which span a large range of exposure times,
are corrected for tilts in the sky illumination (caused by the large f.o.v),
combined with outlier rejection to produce the flatfield and a shutter correction
frame. This frame corrects for the uneven illumination pattern caused by the opening
and closing of the iris shutter used in the SuperWASP cameras. This is a very small
effect ($\lesssim0.1\%$) for our 30\,s exposures but is included for completeness
within the pipeline as it may be used on data where this effect is more pronounced.
Master calibration files from previous nights are then combined with weights that
decay exponentially over time with a ``half-life'' of 14 days.

Science frames have bad pixels masked, are bias \& dark-subtracted, corrected for
shutter travel time and
flat-fielded using \textsc{kappa} \citep{kappa} and \textsc{figaro} \citep{figaro}
routines. Objects are then detected on the frame using \textsc{SExtractor}
\citep{bertin96} as packaged by Starlink. An automated triangle matching algorithm
is used to match the catalogue of detected CCD objects with an automatically
extracted subset of the Tycho-2 \citep{hog00tycho2} catalogue and derive an
astrometric solution. The 9 co-efficient astrometric fit, which allows for fitting
of the field centre and barrel distortion, typically has an RMS precision of
0.1--0.2 pixels.

The sky level is determined using a quadratic fit with outlier rejection to the
sky background with the stellar sources masked out. Aperture photometry is then
carried out on all objects that are present in the frame or have an entry in the
USNO-B1.0 \citep{monet03usnob} catalogue with a 2nd epoch red magnitude less
than 15.0 within the frame. Objects that are detected within the frame by
\textsc{SExtractor} but are not present within the catalogue, are designated as
``orphans'' and are assigned a \SW\ identifier based on their position and
exported into the FITS binary catalogue which is passed to the photometry code.
This ensures that detected transient objects are included and measured. Fluxes
are then measured in three apertures of radii 2.5, 3.5 and 4.5 pixels and ratios
between these fluxes were used to define a ``blending index'' to aid in
filtering out blended and non-stellar objects.

The frames in each field were then post-processed to correct for primary and
secondary extinction with the frame zero-points tied to a set of local secondary
calibrators for each field. These local calibrators were produced from stars
observed on high quality, dark moon nights with the magnitudes obtained from the
WASP fluxes, transformed through a colour equation relating the instrumental and
Tycho-2 magnitudes. Finally the resulting FITS binary catalogues for each frame
are uploaded to the Atlas DataStore operated by RAL and then ingested by the
SuperWASP Data Archive at University of Leicester.

\begin{table*} 
\centering
\caption{Co-ordinates of field centres surveyed in this work.$^\dagger$ Includes
3 objects detected in more than 1 field.} 
\protect\label{tab:fields} 
\begin{tabular}{ccccccccc} 
\hline 
 RA   	  & Dec       	    & Camera & No. of & No. of & No. of extracted & No. of initial & No. of Filtered & Final no. of\\
\multicolumn{2}{c}{(J2000.0)} & (DAS) no. &  nights & stars  & stars & candidates & candidates & candidates \\
\hline 
17 16 00  & +31 26 00 & 3 & 127 &  40438 &  8656 &  664 &  0 & 0 \\
17 17 00  & +23 26 00 & 4 & 129 &  46860 &  9516 &  732 &  3 & 1 \\ 
17 38 00  & +55 41 00 & 5 & 110 &  41081 &  8201 &  220 &  0 & 0 \\
17 39 00  & +47 23 00 & 2 & 119 &  44388 &  8791 &  410 &  9 & 2 \\
17 41 00  & +40 24 00 & 1 & 103 &  44175 &  9851 &  596 &  6 & 2 \\
17 43 00  & +31 26 00 & 2 & 130 &  51411 & 11681 &  619 &  2 & 1 \\
17 44 00  & +24 27 00 & 1 & 113 &  63467 & 13893 &  653 &  6 & 2 \\
17 44 00  & +39 44 00 & 5 & 122 &  48612 & 11033 &  911 &  1 & 1 \\
17 45 00  & +10 28 00 & 1 &  93 &  98296 & 21164 & 1211 &  8 & 2 \\
17 45 00  & +17 27 00 & 2 & 110 &  84334 & 17818 &  691 &  4 & 1 \\
17 46 00  & +25 45 00 & 5 & 108 &  59452 & 14267 &  656 &  4 & 0 \\
18 14 00  & +17 27 00 & 3 & 110 & 116646 & 24216 & 1321 &  7 & 1 \\
18 15 00  & +09 28 00 & 4 & 109 & 178561 & 26672 & 1163 &  8 & 0 \\
\hline
\multicolumn{4}{r}{Totals} & 917721 & 185759 & 9847 & \ntr$^\dagger$ & 13$^\dagger$ \\
\hline
\end{tabular}
\end{table*}

\begin{figure*}
\def\subfigtopskip{0pt}    
\def\subfigbottomskip{4pt}
\def\subfigcapskip{1pt}
\centering

\begin{tabular}{cc}
\subfigure[SW1745+1028 (93)]{\label{fig:SW1745+1028trancov}
\includegraphics[angle=270,width=8.5cm]{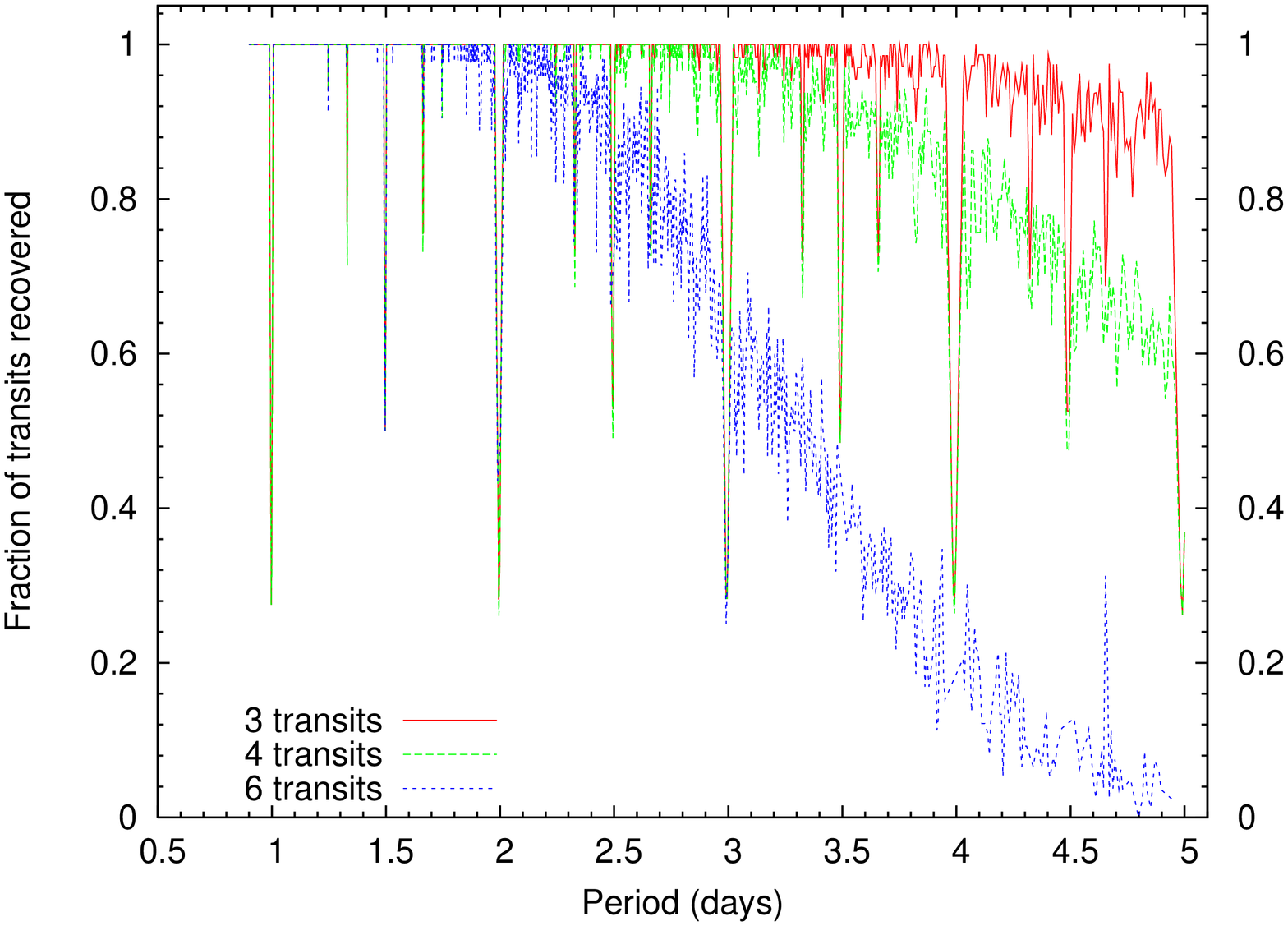}
}
&
\subfigure[SW1745+1727 (110)]{\label{fig:SW1745+1727trancov}
\includegraphics[angle=270,width=8.5cm]{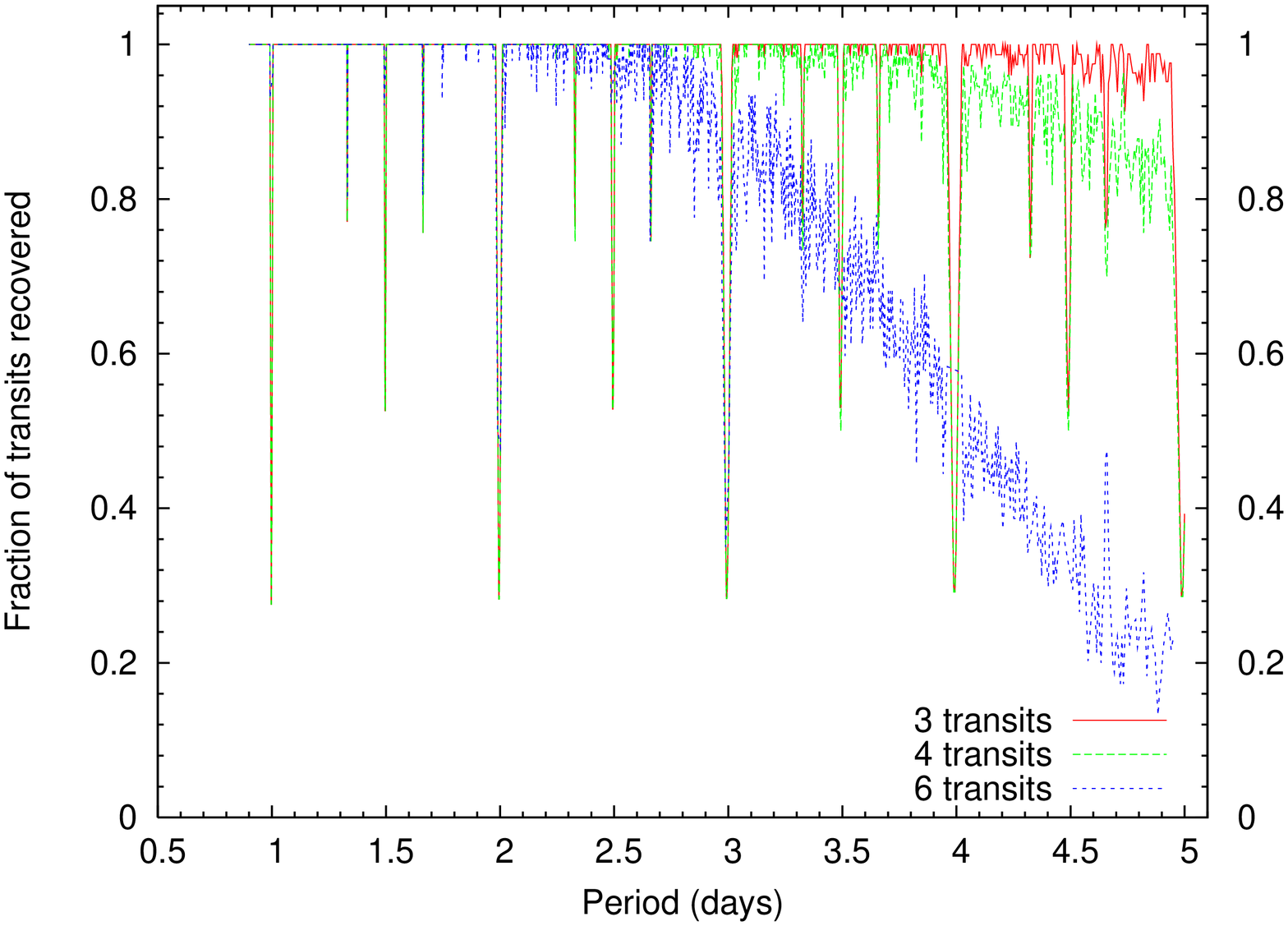}
} \\

\subfigure[SW1739+4723 (119)]{\label{fig:SW1739+4723trancov}
\includegraphics[angle=270,width=8.5cm]{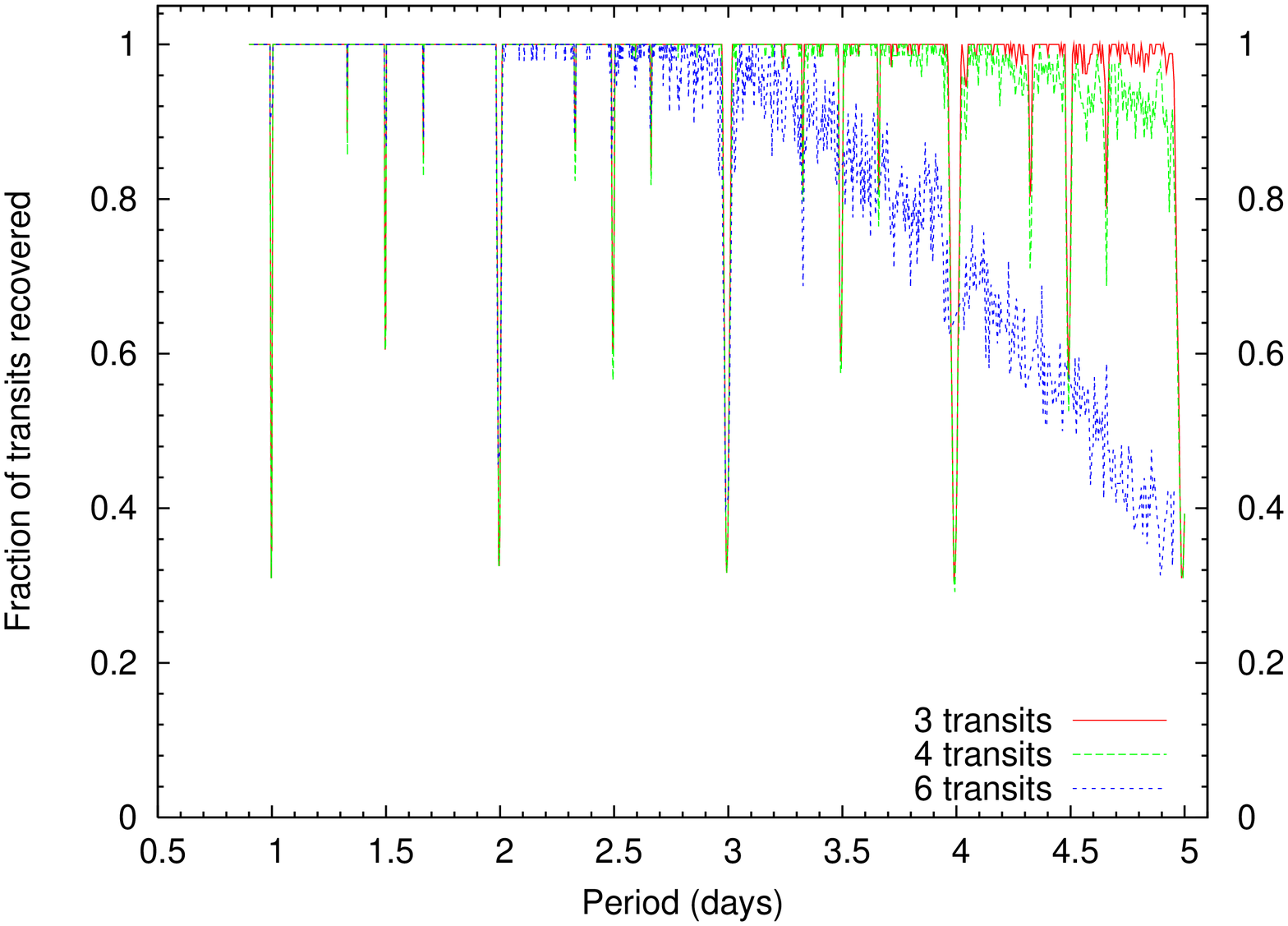}
}
&
\subfigure[SW1743+3126 (130)]{\label{fig:SW1743+3126trancov}
\includegraphics[angle=270,width=8.5cm]{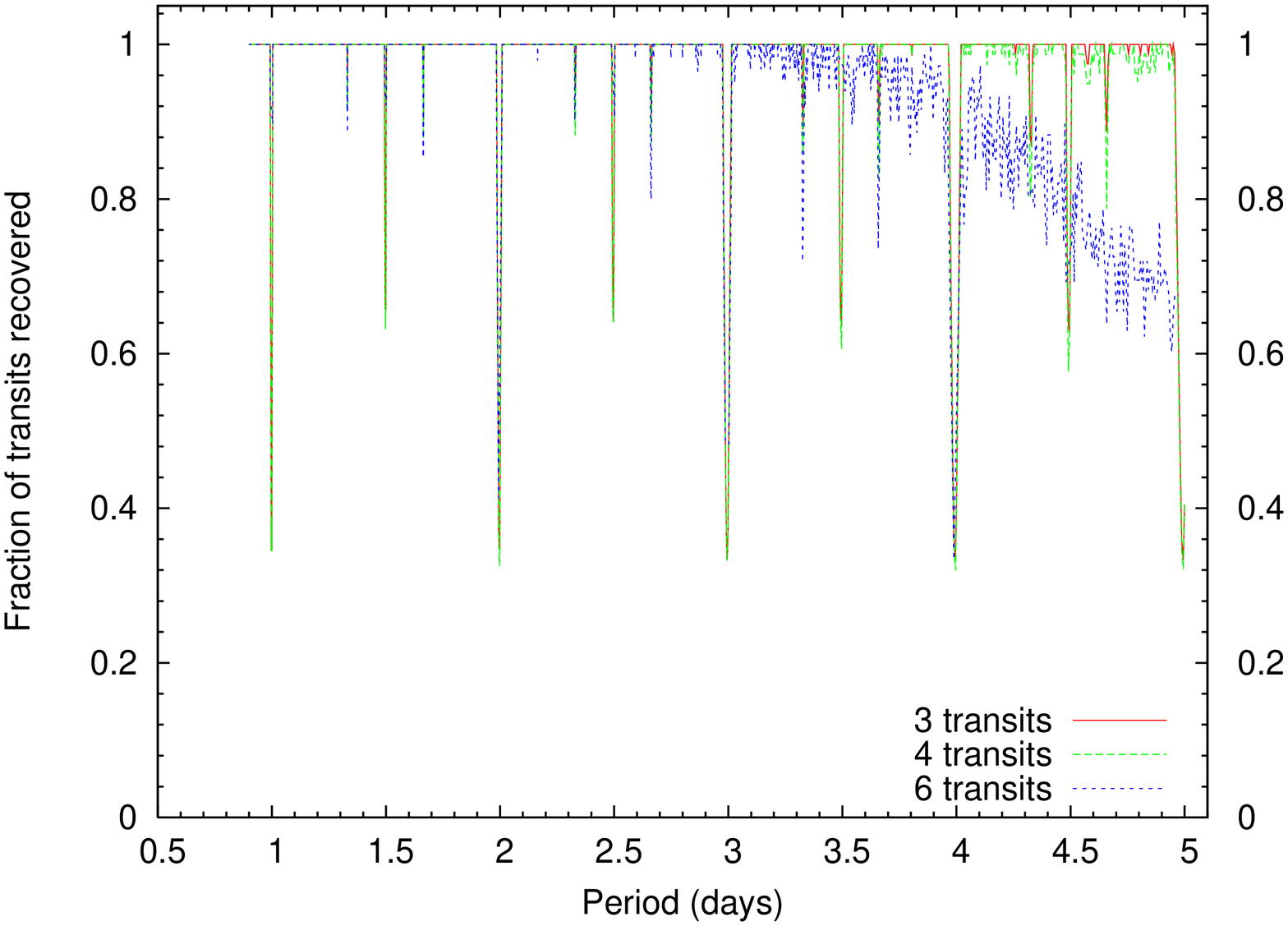}
} \\
\end{tabular}
\caption[]{Fraction of transits recovered for four fields with $\sim90$,110, 120
and 130 nights observed. Results for 3 (top curve) , 4 (middle curve) and 6
(lower curve) detected transits are shown
and the number of nights observed are shown in brackets.}
\label{fig:trancov}
\end{figure*}

\subsection{Field selection \& archive extraction}
\protect\label{sec:archive}

Data for each field were obtained from the SuperWASP Data Archive held at the
University of Leicester. The data were extracted for each field and overlapping
data from other cameras in adjacent fields were rejected. In addition we
required a star to have at least 500 data points on more than 10 nights and WASP
$V\lesssim 13$ in order for it to be included in the transit search. The details
of the 13 fields extracted are given in Table~\ref{tab:fields}. The extraction
process and the requirements on the minimum number of observations of a star
results in a total of $\sim186,000$ extracted stars from a total of
$\sim920,000$.

\section{Selection of candidates}
\protect\label{sec:select}

\subsection{Stage 1 - Transit searching}
\protect\label{sec:transsearch}

The full details of the removal of the systematic errors left in the data after
post-processing and the transit search strategy are given in \citet{cameron06}
but we present a brief summary here.

Although the \SW\  fluxes are referenced to a set of stable standard stars for each
field drawn from Tycho-2 \citep{hog00tycho2}, there are still small systematic
differences in the frame to frame zero-point and colour-dependent terms
introduced by the broad unfiltered bandpass. We remove these systematics using
the \textsc{sysrem} algorithm \citep{tamuz05}.

The transit searching code (\textsc{huntsman}) uses a refined version of the
Box Least Squares (BLS) algorithm \citep{kovacs02} which has been shown by
\citet{tingley03} to be the optimal search method. An initial coarse search
grid is set up over frequencies (defined in terms of the period, $P$ with $0.9
\leq P \leq 5$\, days) and transit epochs ($T_0$) with the transit width ($W$)
calculated at each frequency via Kepler's 3rd law, assuming a stellar mass of
0.9\Msun. The transit depth ($\delta$) and the goodness of fit statistic,
$\chi^2$, are calculated using optimal fitting.

After the coarse grid search, potential transit candidates have to pass
a series of tests designed to weed out false positives. The tests that lead to
rejection as a potential candidate are:
\begin{enumerate}
\item Less than two transits observed
\item Reduced $\chi^2$ of the best fit greater than 3.5
\item Any phase gaps greater than $2.5\times$ the transit width
\item Signal to red noise ratio, $S_{\rmn{red}}<5$
\item The ratio of $\Delta\chi^2$ with a transit model to the \dchs of an
anti-transit model (following \citealt{burke06}) less than 1.5
\end{enumerate}
The signal to red noise ratio ($S_{\rmn{red}}$ - \citealt{cameron06},
\citealt{pont06rednoise}) is the ratio of the best-fit transit depth to the RMS
scatter when binned on the expected transit duration and gives a measure of the
reliability of a transit detection.

Each candidate in the reduced sample that passes the above tests then has the
transit parameters of the five most significant peaks in the periodogram
refined. In this stage the pure box function of \citet{kovacs02} is replaced
with a softened version due to \citet{protopapas05}. This is analytically
differentiable with respect to the key transit parameters ($P, T_0, W,
\delta$), allowing rapid refinement using the Newton-Raphson method.

If a transit passes all these tests then it is accepted as an initial transit
candidate. The numbers of these for each field are shown in the `No. of initial
candidates' column of Table~\ref{tab:fields}. 

One by-product of the transit searching is an investigation of the transit
recovery fraction as a function of period and differing requirements on the
number of transits required. The results of this are shown in
Figure~\ref{fig:trancov} for four sample fields with the number of nights
observed spanning the range shown in Table~\ref{tab:fields}. The figure gives
the probability of at least $N$ transits (for $N=\{3,4,6\}$) being present in
the data as a function of orbital period. Transits are considered to be present
if there are data within the range of phases $\phi<0.1W/P$ or $\phi>1-0.1W/P$
where $W$ is the expected transit width and $P$ is the orbital period (see
Section 3.1 of \cite{cameron06})

\subsection{Stage 2 - Visual inspection}
\protect\label{sec:visinspec}

The transit search described in Section~\ref{sec:transsearch} produced a initial
list of 9847 transit candidates. These were ordered in order of decreasing
signal to red noise and were all visually inspected. The candidates are
identified by their SuperWASP identifiers which are of the form `1SWASP
J\textit{hhmmss.ss}+\textit{ddmmss.s}', with the co-ordinates based on their
position for epoch and equinox J2000.0.

A large fraction ($>50$\%) of the putative transit candidates were caused by
defects in the photometry being folded on the 1 day alias and multiples of it.
False candidates caused by the 1 day alias were found at periods of $\sim2,
\sim3, \sim1.5$ and $\sim 1.33$ days as well as the normal $P\sim1$ day. These
were rapidly eliminated on casual examination of the light curves and
periodograms during the visual inspection process.

During the visual inspection of the transit candidates, any candidates that
showed plausible transit shape and depth, remained flat outside of eclipse and
were not close to an aliased period were recorded to be carried through to the
next stage. To quantify this somewhat subjective process, we developed the
following 4-digit coding scheme:
\begin{itemize}
\item 1st Digit: Shape and visibility of the transit.  
\renewcommand{\labelenumi}{\arabic{enumi}.}
  \begin{enumerate}
  \item Clear transit-shaped signal of credible width and depth.
  \item Shallow/noisy but clearly visible transit signal.
  \item Transit barely visible, either very shallow, lost in noise or ill-shaped.
  \item Partial transit or gaps around phase 0 but still showing clear transit
  morphology.
  \item Signs of a dip at phase 0 but no clear in/egress.
  \end{enumerate}
  
\item 2nd Digit: Out-of-transit light curve.
  \begin{enumerate}
  \item Clean and flat, no other variations.
  \item Noisy but flat.  
  \item Signs of ellipsoidal variation or suspected secondary eclipses (includes
  some candidates which have been folded on twice the period).
  \item Shows low-amplitude sinusoidal variation on short timescales, giving a 
  `knotty' appearance (can indicate that the light curve is folded on the wrong
  period).  
  \item Realistic variability of some other form out of transit.  
  \item Multi-level or `jumpy' light curves (can indicate the wrong period or
  photometry artifacts).  
  \end{enumerate}
  
\item 3rd Digit: Distribution of points in the folded light curve.
  \begin{enumerate}
  \item Smoothly sampled with a similar density of points throughout.
  \item Some minor regions with slightly lower density of points, retaining a
  clear signal. 
  \item Significant clumpiness of data points (can indicate a pathological period).
  \end{enumerate}
  
\item 4th Digit: Credibility of determined period.
  \begin{enumerate}
  \item No reason to doubt measured period, clear peak in $\Delta \chi^{2}$
  periodogram.  
  \item Period gives a secure signal visible in the folded light curve, but
  peak lies close to a known alias.  Sometimes associated with gaps in the folded
  light curve.  
  \item Signal visible in folded light curve but period is a known alias or peak
  lies at a commonly-occurring frequency.  
  \item Light curve suggests that the measured period is wrong.  
  \end{enumerate}
\end{itemize}
\renewcommand{\labelenumi}{\roman}

These codes are shown in the last column of Table~\ref{tab:newinitialcands1}. We
emphasise that these codes are not designed as an ``algorithmic'' means of
eliminating candidates with a certain code, merely a way to attempt to quantify
the subjective visual assessment.

\subsection{Stage 3 - Filtering}
\protect\label{sec:filter}

After the candidates for a field have been identified through visual inspection,
filtering is performed so that a valid candidate is required to have:

\begin{itemize}
\item Signal to red noise ratio ($S_{\rmn{red}})\geq8$
\item Period greater than 1.05 days.
\item More than 3 transits observed.
\item Transit to anti-transit ratio (\dchs/\dchs$_- \geq 2.0$)
\item Signal-to-noise of the ellipsoidal variation (S/N$_{ellip})<8$ \\(based on a
cosine fit to the out of transit data - see \citealt{cameron06} for more details)
\end{itemize}

Following this filtering, the initial candidates were re-sorted into RA order
and any groups of objects with very similar transit parameters and \SW\
identifiers, indicating close proximity on the sky, were investigated using VSI
(see next section). In all of these cases the postage stamp images indicated the
groups of objects were within the same aperture and therefore almost
certainly blended together and they were removed from the list.

In a few cases, candidates have been carried forward to the next stage of
filtering with values of these parameters outside the above ranges. This has
generally been in the cases where a value is very close to the cutoff or there
has been no evidence for the significant ellipsoidal variation suggested by the
S/N$_{ellip}$ value and we have erred on the side of inclusion.

The candidates surviving this filtering are shown in
Table~\ref{tab:newinitialcands1} along with other information such as the signal
to red noise ratio ($S_{\rmn{red}}$) period ($P$), duration and depth ($\delta$)
of the transit, the delta chi-squared of the model (\dchs), number of transits
($\rmn{N_{tr}}$), the signal-to-noise of the ellipsoidal variation
(S/N$_{ellip}$) and the transit to anti-transit ratio (\dchs/\dchs$_-$).

\begin{table*}
\centering
\caption{Low-amplitude candidate extrasolar planets.}
\protect\label{tab:newinitialcands1}
\begin{tabular}{lcccccccccc}
\hline
Ident. (1SWASP +) & Period & $\delta$ & Duration & Epoch & $\rmn{N}_{tr}$ & \dchs & \dchs/ & $\rmn{S/N}_{ellip}$ & $S_{red}$ & LC Code \\ 
	      & (days) & (mag) & (hrs)       & (2450000+) &     &  	     & \dchs$_-$ &   &		       & \\
\hline
\multicolumn{11}{c}{Field SW1717+2326} \\
\hline
\textbf{J165949.13+265346.1} & \textbf{2.683010} & \textbf{0.0209} &  \textbf{1.824} & \textbf{3128.1597} & \textbf{13} &  \textbf{960.5991} & \textbf{4.8709} &  \textbf{9.717} & \textbf{12.329} & \textbf{1111} \\ 
J170319.50+271317.8 & 1.527922 & 0.0528 &  1.656 & 3128.1069 & 22 &  949.4885 & 4.0610 &  5.519 & 12.647 & 2211 \\ 
J172515.66+234853.9 & 1.457706 & 0.0361 &  1.440 & 3128.1826 & 23 &  553.5663 & 4.0588 &  3.733 & 10.901 & 2311 \\ 
\hline
\multicolumn{11}{c}{Field SW1739+4723} \\
\hline
J172117.67+441817.8 & 1.941607 & 0.0336 &  3.912 & 3137.6143 & 19 &  813.7049 & 3.9546 &  7.026 & 12.459 & 1212 \\ 
J172302.03+472043.0 & 4.458795 & 0.0380 &  3.264 & 3136.5464 &  9 &  717.0023 & 2.9865 &  2.683 & 12.834 & 1211 \\ 
J172336.03+462044.5 & 1.162678 & 0.0153 &  2.232 & 3138.4099 & 29 &  236.8270 & 3.5019 &  4.606 &  9.860 & 3211 \\ 
\textbf{J172549.13+502206.4} & \textbf{4.542614} & \textbf{0.0345} &  \textbf{1.536} & \textbf{3137.4856} &  \textbf{8} & \textbf{569.7906} & \textbf{2.8470} &  \textbf{6.495} & \textbf{11.941} & \textbf{1211} \\ 
(J172826.46+471208.4)$^c$ & 3.405044 & 0.0217 &  2.496 & 3139.1104 &  7 &  334.1523 & 1.1349 &  1.889 & 11.967 & 4121 \\ 
J173253.52+435009.9 & 4.557160 & 0.0287 &  4.488 & 3139.5488 &  8 & 1263.1642 & 3.7183 &  1.259 & 11.549 & 2211 \\ 
J173428.91+471225.3 & 4.304363 & 0.0334 &  2.328 & 3136.9143 &  6 &  139.1181 & 3.2450 &  0.773 &  9.994 & 2222 \\ 
\textbf{J173748.98+471348.7} & \textbf{3.337786} & \textbf{0.0105} &  \textbf{3.456} & \textbf{3139.1697} & \textbf{10} & \textbf{390.9406} & \textbf{4.3549} &  \textbf{1.028} &  \textbf{9.895} & \textbf{2111} \\ 
J174619.33+450103.3 & 2.403553 & 0.0295 &  3.336 & 3138.1704 & 15 & 1442.1517 & 3.6050 &  4.681 & 13.163 & 1311 \\ 
J180518.31+460504.9 & 3.037774 & 0.0506 &  2.352 & 3138.4680 &  5 & 1142.2657 & 14.8579 & 3.273 & 10.120 & 5222 \\ 
\hline
\multicolumn{11}{c}{Field SW1741+4024} \\
\hline
J174116.85+383706.2 & 4.242378 & 0.0134 &  2.832 & 3138.4702 &  6 &  149.0061 & 3.3432 &  0.334 & 11.368 & 2122 \\ 
\textbf{J174118.30+383656.3} & \textbf{4.245098} & \textbf{0.0128} &  \textbf{3.264} & \textbf{3138.4246} &  \textbf{8} &  \textbf{137.7380} & \textbf{4.8031} &  \textbf{1.416} & \textbf{12.554} & \textbf{1211} \\ 
J174959.05+370928.8 & 2.530897 & 0.0302 &  5.472 & 3137.0903 & 16 &  729.3705 & 17.1295 &  7.504 & 14.313 & 1322 \\ 
J175138.04+381027.5 & 1.543934 & 0.0544 &  2.520 & 3139.4077 & 17 & 6584.2744 & 14.1000 &  3.550 & 14.826 & 1112 \\ 
J175207.01+373246.3 & 1.306420 & 0.0198 &  1.032 & 3138.3350 & 20 &  195.1881 &  1.4261 &  0.078 & 11.228 & 2211 \\ 
\textbf{J175856.34+421950.9} & \textbf{3.256700} & \textbf{0.0221} &  \textbf{2.088} & \textbf{3137.7993} &  \textbf{8} &  \textbf{430.9857} & \textbf{9.0792} &  \textbf{0.849} & \textbf{14.039} & \textbf{1211} \\ 
\hline
\multicolumn{11}{c}{Field SW1743+3126} \\
\hline
J174343.15+340306.5 & 2.322198 & 0.0443 &  4.104 & 3128.0608 & 19 & 2023.3019 & 5.3088 &  0.326 & 10.674 & 1231 \\ 
(J174645.84+333411.9)$^c$ & 1.571636 & 0.0383 &  1.968 & 3127.0706 & 22 & 10409.9561 & 52.7193 & 18.634 & 24.720 & 1111 \\ 
\textbf{J175401.58+322112.6} & \textbf{1.949258} & \textbf{0.0136} &  \textbf{1.992} & \textbf{3127.8124} & \textbf{20} &  \textbf{225.32}  & \textbf{3.0137} &  \textbf{2.911} &  \textbf{9.440} & \textbf{2212} \\
\hline
\multicolumn{11}{c}{Field SW1744+2427} \\
\hline
(J173403.61+280145.1)$^c$ & 4.62676  & 0.0568 &  0.192 & 3126.4275 &  6 &  230.8974 & 4.3579 &  1.356 & 15.863 & 3123 \\
J173508.25+232123.9 & 2.610171 & 0.0438 &  2.088 & 3126.0015 & 11 & 2530.1182 & 2.3779 &  0.218 & 15.612 & 1111 \\ 
J174221.53+271435.2 & 2.104947 & 0.0253 &  2.640 & 3126.7256 & 14 &  229.8633 & 4.5496 &  1.893 & 12.803 & 4212 \\ 
J175143.72+205953.9 & 3.069992 & 0.0145 &  3.240 & 3126.0132 & 10 &  297.5705 & 2.1322 &  2.271 & 12.900 & 1211 \\ 
J175236.10+273225.3 & 1.905583 & 0.0298 &  3.216 & 3127.1951 & 21 & 4562.8867 & 12.3990 & 12.173 & 18.554 & 1112 \\ 
\textbf{J175620.84+253625.7} & \textbf{4.415010} & \textbf{0.0340} &  \textbf{2.568} & \textbf{3124.5637} &  \textbf{8} &  \textbf{974.5581} & \textbf{15.1208} &  \textbf{2.012} & \textbf{15.219} & \textbf{1211} \\ 
\textbf{J180010.55+214510.2} & \textbf{3.434074} & \textbf{0.0557} &  \textbf{2.184} & \textbf{3125.6943} &  \textbf{8} &  \textbf{552.4845} & \textbf{7.3147} &  \textbf{0.537} & \textbf{16.020} & \textbf{1211} \\ 
\hline
\multicolumn{11}{c}{Field SW1744+3944} \\
\hline
\textbf{J175856.34+421950.9} & \textbf{3.258676} & \textbf{0.0301} &  \textbf{2.184} & \textbf{3127.9993} & \textbf{10} &  \textbf{431.7925} & \textbf{2.4261} &  \textbf{1.041} & \textbf{12.111} & \textbf{1212} \\ 
\hline
\multicolumn{11}{c}{Field SW1745+1028} \\
\hline
J172917.65+065655.0 & 0.931937 & 0.0217 &  1.464 & 3150.7373 & 23 &  445.3489 & 4.5051 &  0.581 & 12.132 & 2212 \\ 
J173238.84+104059.9 & 2.283212 & 0.0189 &  2.280 & 3150.2688 & 10 &  191.4931 & 3.8177 &  1.191 & 16.949 & 4111 \\ 
J173631.20+133442.9 & 1.609477 & 0.0173 &  1.368 & 3151.2756 & 14 &  225.8527 & 5.1731 &  1.606 & 12.332 & 2211 \\ 
\textbf{J174058.24+062638.1} & \textbf{4.804517} & \textbf{0.0168} &  \textbf{4.560} & \textbf{3146.9646} &  \textbf{7} &  \textbf{286.7106} & \textbf{2.5973} &  \textbf{0.845} &  \textbf{9.583} & \textbf{3212} \\ 
J174155.92+081459.1 & 1.228380 & 0.0363 &  3.216 & 3150.6472 & 22 & 1211.9445 & 6.8155 &  3.168 & 13.418 & 2211 \\ 
J174222.47+101901.5 & 3.344754 & 0.0168 &  3.576 & 3148.4062 &  7 &  173.1803 & 3.3064 &  4.469 &  8.878 & 2212 \\ 
\textbf{J175511.09+134731.5} & \textbf{2.444503} & \textbf{0.0201} &  \textbf{2.712} & \textbf{3149.3391} & \textbf{10} &  \textbf{355.9681} & \textbf{7.0766} &  \textbf{1.328} & \textbf{14.494} & \textbf{1111} \\ 
J175813.15+095151.2 & 3.746969 & 0.0488 &  1.728 & 3149.6096 &  6 &  524.3446 & 8.5659 &  1.590 & 15.826 & 1221 \\ 
\hline
\multicolumn{11}{c}{Field SW1745+1727} \\
\hline
J174100.71+154714.9 & 2.147224 & 0.0551 &  3.264 & 3150.0034 & 14 & 4797.7295 & 21.8317 & 16.977 & 20.532 & 1314 \\
J174656.28+143841.2 & 4.026640 & 0.0322 &  2.520 & 3147.7732 &  6 &  271.6478 & 2.2670 &  1.491 &  8.816 & 2222 \\
J175143.72+205953.9 & 3.070980 & 0.0149 &  3.504 & 3150.5601 & 13 &  366.4742 & 3.4870 &  0.186 & 12.832 & 1111 \\
\textbf{J175511.09+134731.5} & \textbf{2.443916} & \textbf{0.0242} &  \textbf{2.520} & \textbf{3149.3479} & \textbf{12} &  \textbf{664.5929} & \textbf{4.6498} &  \textbf{5.335} & \textbf{13.218} & \textbf{1111} \\ 
\hline
\multicolumn{10}{l}{$(<Id>)^c$ Not selected as a candidate but has spectroscopic data} \\
\end{tabular}
\end{table*}

\begin{table*}
\centering
\contcaption{Low-amplitude candidate extrasolar planets.}
\protect\label{tab:newinitialcands2}
\begin{tabular}{lcccccccccc}
\hline
Ident. (1SWASP +) & Period & $\delta$ & Duration & Epoch & $\rmn{N}_{tr}$ & \dchs & \dchs/ & $\rmn{S/N}_{ellip}$ & $S_{red}$ & LC Code \\ 
	       & (days) & (mag) & (hrs)    & (2450000+) &     & 	      & \dchs$_-$ &  &		  & \\
\hline
\multicolumn{11}{c}{Field SW1746+2545} \\
\hline
J173822.25+290549.2 & 2.139874 & 0.0633 &  3.792 & 3151.9006 & 10 &  918.5665 & 4.9992 &  7.531 &  9.800 & 1621 \\ 
J174448.71+273630.5 & 1.872146 & 0.0223 &  2.592 & 3152.6758 & 14 & 1343.4296 & 5.2142 &  0.701 & 13.378 & 1214 \\ 
J175236.10+273225.3 & 1.905144 & 0.0328 &  2.928 & 3151.9788 & 18 & 1888.8925 & 4.1447 & 21.394 & 14.569 & 1112 \\ 
J180004.71+255947.6 & 1.238110 & 0.0212 &  2.880 & 3153.4089 & 24 &  313.1419 & 1.5115 &  2.016 & 10.483 & 2211 \\ 
\hline
\multicolumn{11}{c}{Field SW1814+1727} \\
\hline
\textbf{J175914.99+213803.9} & \textbf{4.552594} & \textbf{0.0302} &  \textbf{4.224} & \textbf{3148.7092} &  \textbf{4} &  \textbf{184.8400} & \textbf{2.0828} &  \textbf{0.090} &  \textbf{8.105} & \textbf{2221} \\ 
J181022.15+172132.3 & 1.055949 & 0.0219 &  2.424 & 3150.6555 & 28 &  639.9628 & 7.8049 &  6.630  & 15.093 & 1112 \\ 
J181113.13+141441.9 & 3.281877 & 0.0287 &  3.360 & 3148.8628 & 12 &  370.3304 & 7.4041 &  0.700 & 12.922 & 2211 \\
J182330.47+160218.4 & 1.201000 & 0.0528 &  2.232 & 3151.4124 & 22 & 13563.5039 & 60.1122 &  8.092 & 21.624 & 1111 \\ 
J182428.52+160346.2 & 4.258705 & 0.0476 &  2.952 & 3148.2244 &  6 & 1289.5090 & 2.2782 &  3.331 & 14.273 & 1411 \\ 
J182851.64+200727.2 & 2.381931 & 0.0252 &  2.112 & 3150.8325 & 13 &  351.3000 & 4.1220 &  3.022 & 13.246 & 2211 \\ 
J182957.77+174455.2 & 1.178835 & 0.0252 &  2.856 & 3150.5652 & 23 &  965.1992 & 4.3350 &  9.531 & 13.331 & 2211 \\ 
\hline
\multicolumn{11}{c}{Field SW1815+0928} \\
\hline
J175913.94+132849.4 & 2.980343 & 0.0489 &  3.024 & 3150.7480 &  9 & 4167.3213 & 18.6379 &  5.496 & 12.770 & 1133 \\ 
J180202.52+065737.9 & 1.760327 & 0.0246 &  1.968 & 3150.7300 &  9 &  603.1787 & 7.1428 &  1.815 & 12.902 & 1321 \\ 
J181222.90+100032.6 & 1.405853 & 0.0434 &  2.232 & 3151.4324 & 14 & 1102.3527 & 15.8264 &  1.541 & 16.583 & 1111 \\ 
J181858.42+103550.1 & 2.464850 & 0.0124 &  2.760 & 3151.1021 & 10 &  206.5994 & 2.7807 &  0.529 & 12.086 & 2122 \\ 
J182127.51+094038.2 & 1.832481 & 0.0145 &  2.784 & 3150.8479 & 11 &  215.1272 & 2.1370 &  1.357 &  8.855 & 2111 \\ 
J182317.92+063936.2 & 1.854633 & 0.0300 &  2.160 & 3151.1941 & 13 &  262.3272 & 7.5657 &  1.000 & 12.530 & 3212 \\ 
J182543.36+122925.0 & 1.390569 & 0.0539 &  1.680 & 3150.7791 & 15 & 3254.0181 & 9.1157 &  2.034 & 14.618 & 1111 \\ 
J182741.05+082414.0 & 4.875473 & 0.0473 &  4.152 & 3147.1956 &  6 &  441.6472 & 8.8947 &  3.500 & 11.275 & 2222 \\ 
\hline
\end{tabular}
\end{table*}

The visual inspection process reduced the 9847 initial transit candidates to 199
and the filtering process then further reduced the number of candidates to
\ntr, including 3 candidates which were detected in more than one field (a total
of \ntru\ unique objects). The number of these objects per field is shown in the
`No. of filtered candidates' column of Table~\ref{tab:fields}.

\subsection{Stage 4 - Additional candidate information}
\protect\label{sec:candfilt}

Once filtered transit candidates have been identified from the light curves, we
make use of Variable Star Investigator (VSI) to provide additional information
on the transit candidates. VSI was written by one of us (DMW) to query large
numbers of astrometric catalogues (USNO-B1.0, Tycho-2, 2MASS, UCAC-2, PPM),
variable object catalogues (ROSAT, CCDM, GCVS), and the image servers (DSS,
2MASS (\citealt{skrutskie062mass})) using the extracted data and atlas images
to find blended objects and nearby companions.

VSI also makes use of transit depth, width and period from \textsc{huntsman},
combined with the colours and radius estimation (\citealt{cox00};
\citealt{ammons06}; \citealt{gray92}) for the extracted stars and the expression from
\citet{tingley05}:
\[
R_p \simeq  R_{*} \sqrt{\frac{\delta}{1.3}},
\]
to estimate the planetary radius.

The factor of 1.3 in the above equation is intended to account for
limb-darkening effects. It is derived from Monte Carlo simulations and is
strictly only valid for observations in the $I$ band but the difference between
$I$ and our unfiltered wide bandpass is minimal given the errors on the stellar
radius when estimated from broadband colours. 

We also use the $\eta_p$ exoplanet diagnostic from \citet{tingley05} which is
defined as:
\[
\eta_p \equiv \frac{D_{obs}}{D}
\]
where $D_{obs}$ is the observed transit duration and $D$ is the
theoretical transit duration. The theoretical duration can be expanded in terms of
the period, planetary radius $R_p$ and the transit depth (see \citet{tingley05}
for further details) enabling $\eta_p$ to be calculated from the supplied
transit parameters and the information extracted from the catalogues. Values
close to one indicate the observed and theoretical durations agree well and the
candidate is more likely to be genuine.

In Table~\ref{tab:filtercands1} we give the $V_{SW}-K$
colour, the $J-H$ and $H-K$ colours from 2MASS, the inferred stellar radius
($R_*$ in solar radii) and planetary radius ($R_p$ in Jupiter radii), the
exoplanet diagnostic ($\eta_p$), the number of brighter ($N_\rmn{br}$) and
$<5$\,mag fainter ($N_\rmn{faint}$) objects within the 48\arcsec\ aperture and a
series of letter codes for the plausibility of the planetary radius, exoplanet
diagnostic and the degree and severity of the blending within the 48\arcsec\ 
aperture. 

The letter codes are as follows:
\begin{itemize}
\item Planetary radius (R): Values range from A ($R_p<1.6$), B ($1.6\geq R_p
> 1.75$) to C ($R_p \geq1.75$))
\item Exoplanet diagnostic (Eta): Values are encoded as follows A ($0.5\geq
\eta_p \geq 1.5$), B ($\eta_p<0.5$), C ($\eta_p > 1.5$)
\item Blending (B): Codes are A (no blends), B (1 or 2 objects less
than 5 mag fainter in aperture), C ($>2$ fainter objects in aperture), D
(brighter object in aperture)
\end{itemize}

It should be noted that candidates having large inferred planetary radii (codes
B and C) could well be interesting in their own right e.g. brown dwarf or late M
companions.

\begin{table*}
\centering
\caption{Radius and blending results}
\protect\label{tab:filtercands1}
\begin{tabular}{lcccccccccccc}
\hline
Ident. (1SWASP +) & $V_{SW}$ & $V_{SW}-K$ & $J-H$ & $H-K$ & $R_*$ & $R_p$ & $\eta_p$ & ${N^a_\rmn{br}}$ & ${N^b_\rmn{faint}}$ & \multicolumn{3}{c}{Codes$^\ddagger$}\\ 
		 & (mag) & (mag)      & (mag) & (mag) & (\Rsun) & (\Rj) & 	&   & 	& R & Eta & Blend\\
\hline
\multicolumn{13}{c}{Field SW1717+2326} \\
\hline
\textbf{J165949.13+265346.1} & \textbf{10.951} & \textbf{1.50} & \textbf{0.22} & \textbf{0.08} & \textbf{1.08} & \textbf{1.33} & \textbf{0.61} & \textbf{ 0} & \textbf{ 0} & \textbf{A} & \textbf{A} & \textbf{A} \\ 
J170319.50+271317.8 & 12.944 & 0.67 & 0.18 & 0.06 & 1.65 & 3.23 & 0.49 &  1 &  3 & C & B & D \\ 
J172515.66+234853.9 & 12.707 & 0.01 & 0.32 & 0.03 & 2.02 & 3.27 & 0.40 &  1 &  9 & C & B & D \\ 
\hline
\multicolumn{13}{c}{Field SW1739+4723} \\
\hline
J172117.67+441817.8 & 12.771 &  1.50 & 0.29 & 0.08 & 1.08 & 1.69 & 1.42 &  0 &  0 & B & A & A \\ 
J172302.03+472043.0 & 12.575 &  1.39 & 0.21 & 0.06 & 1.16 & 1.93 & 0.86 &  0 &  0 & C & A & A \\ 
J172336.03+462044.5 & 12.684 &  2.95 & 0.64 & 0.12 & 0.68 & 0.72 & 1.32 &  0 &  6 & A & A & C \\ 
\textbf{J172549.13+502206.4} & \textbf{12.033} & \textbf{ 1.70} & \textbf{0.30} & \textbf{0.09} & \textbf{0.97} & \textbf{1.54} & \textbf{0.45} & \textbf{ 0} & \textbf{ 1} & \textbf{A} & \textbf{B} & \textbf{B/A} \\ 
J172824.17+482152.7 & 10.548 &  1.09 & 0.17 & 0.05 & 1.37 & 2.11 & 0.45 &  0 &  0 & C & B & A \\ 
(J172826.46+471208.4)$^c$ & 11.528 & 2.05 & 0.37 & 0.09 & 0.82 & 1.03 & 0.91 &  0 &  0 & A & A & A \\
J173253.52+435009.9 & 11.345 &  2.38 & 0.55 & 0.11 & 0.74 & 1.07 & 1.55 &  0 &  0 & A & C & A \\ 
J173428.91+471225.3 & 12.809 &  N/A  & 0.32 & 0.01 & 1.02$^\dagger$ & 1.59$^\dagger$ & 0.67$^\dagger$ &  4 & 11 & A & A & D \\ 
\textbf{J173748.98+471348.7} & \textbf{11.442} & \textbf{ 1.62} & \textbf{0.33} & \textbf{0.06} & \textbf{1.01} & \textbf{0.88} & \textbf{1.16} & \textbf{ 0} & \textbf{ 1} & \textbf{A} & \textbf{A} & \textbf{B} \\ 
J174619.33+450103.3 & 12.087 &  2.13 & 0.53 & 0.10 & 0.80 & 1.17 & 1.36 &  0 &  4 & A & A & C \\ 
J180518.31+460504.9 & 11.510 &  2.99 & 0.62 & 0.15 & 0.67 & 1.29 & 0.94 &  0 &  3 & A & A & C \\ 
\hline
\multicolumn{13}{c}{Field SW1741+4024} \\
\hline
J174116.85+383706.2 & 11.572 &  N/A  & 0.28 & 0.05 & 1.10$^\dagger$  & 1.09$^\dagger$ & 0.83$^\dagger$ &  1 &  7 & A & A & D \\ 
\textbf{J174118.30+383656.3} & \textbf{11.447} & \textbf{ 1.31} & \textbf{0.22} & \textbf{0.06} & \textbf{1.21} & \textbf{1.17} & \textbf{0.90} & \textbf{ 0} & \textbf{ 1} & \textbf{A} & \textbf{A} & \textbf{B} \\ 
J174959.05+370928.8 & 12.663 &  2.56 & 0.55 & 0.10 & 0.72 & 1.07 & 2.32 &  0 &  4 & A & C & C \\ 
J175138.04+381027.5 & 11.920 &  N/A  & 0.30 & -0.00 & 1.06$^\dagger$ & 2.11$^\dagger$ & 0.96$^\dagger$  &  5 & 11 & C & A & D \\ 
J175207.01+373246.3 & 12.414 &  1.81 & 0.36 & 0.05 & 0.92 & 1.10 & 0.49 &  0 &  2 & A & B & B \\ 
\textbf{J175856.34+421950.9} & \textbf{11.619} & \textbf{ 1.54} & \textbf{0.28} & \textbf{0.06} & \textbf{1.06} & \textbf{1.34} & \textbf{0.66} & \textbf{ 0} & \textbf{ 1} & \textbf{A} & \textbf{A} & \textbf{B} \\ 
\hline
\multicolumn{13}{c}{Field SW1743+3126} \\
\hline
J174343.15+340306.5 & 12.587 &  1.71 & 0.33 & 0.09 & 0.96 & 1.72 & 1.48 &  0 &  2 & B & A & B \\ 
(J174645.84+333411.9)$^c$ & 10.904 &  1.38 & 0.25 & 0.04 & 1.16 & 1.94 & 0.73 &  0 &  0 & C & A & A \\ 
\textbf{J175401.58+322112.6} & \textbf{12.516} & \textbf{1.82} & \textbf{0.28} & \textbf{0.08} & \textbf{0.91} & \textbf{0.91} & \textbf{0.84} & \textbf{ 0} & \textbf{ 0} & \textbf{A} & \textbf{A} & \textbf{A} \\
\hline
\multicolumn{13}{c}{Field SW1744+2427} \\
\hline
(J173403.61+280145.1)$^c$ & 11.421 & 1.42 & 0.23 & 0.08 & 1.14 & 2.32 & 0.05 & 0 & 0 & C & B & A \\
J173508.25+232123.9 & 11.876 &  1.57 & 0.22 & 0.07 & 1.04 & 1.86 & 0.69 &  0 &  0 & C & A & A \\ 
J174221.53+271435.2 & 12.985 &  1.36 & 0.44 & 0.01 & 1.18 & 1.60 & 0.90 &  0 &  2 & B & A & B \\ 
J175143.72+205953.9 & 11.823 &  2.84 & 0.51 & 0.17 & 0.69 & 0.71 & 1.38 &  0 &  2 & A & A & B \\ 
J175236.10+273225.3 & 11.221 &  2.35 & 0.56 & 0.13 & 0.75 & 1.10 & 1.47 &  0 &  9 & A & A & C \\ 
\textbf{J175620.84+253625.7} & \textbf{12.229} & \textbf{ 1.66} & \textbf{0.26} & \textbf{0.04} & \textbf{0.99} & \textbf{1.56} & \textbf{0.75} & \textbf{ 0} & \textbf{ 2} & \textbf{A} & \textbf{A} & \textbf{B/A} \\ 
\textbf{J180010.55+214510.2} & \textbf{12.569} & \textbf{ 2.86} & \textbf{0.52} & \textbf{0.13} & \textbf{0.69} & \textbf{1.39} & \textbf{0.82} & \textbf{ 0} & \textbf{ 4} & \textbf{A} & \textbf{A} & \textbf{C/A} \\ 
\hline
\multicolumn{13}{c}{Field SW1744+3944} \\
\hline
\textbf{J175856.34+421950.9} & \textbf{11.619} & \textbf{ 1.54} & \textbf{0.28} & \textbf{0.06} & \textbf{1.06} & \textbf{1.57} & \textbf{0.68} & \textbf{ 0} & \textbf{ 1} & \textbf{A} & \textbf{A} & \textbf{B/A} \\ 
\hline
\multicolumn{13}{c}{Field SW1745+1028} \\
\hline
J172917.65+065655.0 & 12.288 &  2.47 & 0.58 & 0.15 & 0.73 & 0.92 & 0.88 &  0 &  7 & A & A & C \\ 
J173238.84+104059.9 & 11.357 &  1.12 & 0.25 & 0.02 & 1.35 & 1.58 & 0.71 &  0 &  8 & B & A & C \\ 
J173631.20+133442.9 & 12.004 &  1.61 & 0.36 & 0.05 & 1.02 & 1.14 & 0.57 &  0 &  4 & A & A & C \\ 
\textbf{J174058.24+062638.1} & \textbf{11.745} & \textbf{ 2.05} & \textbf{0.37} & \textbf{0.07} & \textbf{0.82} & \textbf{0.91} & \textbf{1.50} & \textbf{ 0} & \textbf{ 2} & \textbf{A} & \textbf{A} & \textbf{B} \\ 
J174155.92+081459.1 & 12.614 &  2.30 & 0.62 & 0.16 & 0.76 & 1.24 & 1.66 &  0 &  6 & A & C & C \\ 
J174222.47+101901.5 & 12.538 &  1.38 & 0.60 & 0.13 & 1.16 & 1.28 & 1.09 &  1 & 12 & A & A & D \\ 
\textbf{J175511.09+134731.5} & \textbf{11.565} & \textbf{ 1.37} & \textbf{0.25} & \textbf{0.05} & \textbf{1.17} & \textbf{1.42} & \textbf{0.90} & \textbf{ 0} & \textbf{ 0} & \textbf{A} & \textbf{A} & \textbf{A} \\ 
J175813.15+095151.2 & 12.587 &  1.87 & 0.29 & 0.09 & 0.89 & 1.68 & 0.55 &  0 & 10 & B & A & C \\ 
\hline
\multicolumn{13}{l}{$^a$ Number of brighter objects within aperture. $^b$ No. of objects less than 5 mags fainter within aperture.} \\ 
\multicolumn{13}{l}{$(<Id>)^c$ Not selected as a candidate but has spectroscopic data.}\\
\multicolumn{13}{l}{$^\dagger$ Based on $J-H$ colour not a $V_{SW}-K$ colour. }\\
\multicolumn{13}{l}{$^\ddagger$ Codes: R=Planet radius (A=$R_p<1.6$, B=$1.6\geq R_p> 1.75$, C=$R_p \geq1.75$),} \\
\multicolumn{13}{l}{Eta=$\eta_p$ (A=$0.5\geq \eta_p \geq 1.5$, B=$\eta_p<0.5$, C=$\eta_p > 1.5$),} \\
\multicolumn{13}{l}{B=Blending (A=OK, B=1 or 2 fainter objs in aperture, C=$>2$ fainter objs in aperture, D=brighter obj in aperture)}\\
\end{tabular}
\end{table*}

\begin{table*}
\centering
\contcaption{Radius and blending results}
\protect\label{tab:filtercands2}
\begin{tabular}{lcccccccccccc}
\hline
Ident. (1SWASP +) & $V_{SW}$ & $V_{SW}-K$ & $J-H$ & $H-K$ & $R_*$ & $R_p$ & $\eta_p$ & ${N^a_\rmn{br}}$ & ${N^b_\rmn{faint}}$ & \multicolumn{3}{c}{Codes$^\ddagger$}\\ 
		 & (mag) & (mag)      & (mag) & (mag) & (\Rsun) & (\Rj) & 	&   & 	& R & Eta & Blend\\
\hline
\multicolumn{13}{c}{Field SW1745+1727} \\
\hline
J174100.71+154714.9 & 11.653 &  1.23 & 0.21 & 0.05 & 1.27 & 2.54 & 1.01 &  0 &  2 & C & A & B \\ 
J174656.28+143841.2 & 12.263 &  0.50 & 0.36 & 0.10 & 1.85 & 2.83 & 0.53 &  3 &  7 & C & A & D \\ 
J175143.72+205953.9 & 11.823 &  2.84 & 0.51 & 0.17 & 0.69 & 0.72 & 1.49 &  0 &  2 & A & A & B \\ 
\textbf{J175511.09+134731.5} & \textbf{11.565} & \textbf{ 1.37} & \textbf{0.25} & \textbf{0.05} & \textbf{1.17} & \textbf{1.55} & \textbf{0.83} & \textbf{ 0} & \textbf{ 0} & \textbf{A} & \textbf{A} & \textbf{A} \\ 
\hline
\multicolumn{13}{c}{Field SW1746+2545} \\
\hline
J173822.25+290549.2 & 12.449 &  0.95 & 0.30 & 0.02 & 1.46 & 3.13 & 1.07 &  1 &  6 & C & A & D \\ 
J174448.71+273630.5 & 10.792 &  1.19 & 0.23 & 0.03 & 1.30 & 1.66 & 0.88 &  0 &  1 & B & A & B \\ 
J175236.10+273225.3 & 11.221 &  2.35 & 0.56 & 0.13 & 0.75 & 1.16 & 1.33 &  0 &  9 & A & A & C \\ 
J180004.71+255947.6 & 12.615 &  N/A  & 0.07 & 0.04 & 1.88$^\dagger$ & 2.34$^\dagger$ & 0.91$^\dagger$ &  1 & 12 & C & A & D \\ 
\hline
\multicolumn{13}{c}{Field SW1814+1727} \\
\hline
\textbf{J175914.99+213803.9} & 12.585 &  1.86 & 0.30 & 0.08 & 0.89 & 1.32 & 1.30 &  0 &  4 & A & A & C \\ 
J181022.15+172132.3 & 12.661 &  1.61 & 0.28 & 0.12 & 1.02 & 1.29 & 1.15 &  0 & 18 & A & A & C \\ 
J181113.13+141441.9 & 12.629 &  1.72 & 0.27 & 0.07 & 0.96 & 1.39 & 1.11 &  0 &  3 & A & A & C \\ 
J182330.47+160218.4 & 11.066 &  1.17 & 0.18 & 0.10 & 1.31 & 2.57 & 0.82 &  0 &  2 & C & A & B \\ 
J182428.52+160346.2 & 11.788 &  0.75 & 0.09 & 0.08 & 1.55 & 2.89 & 0.65 &  0 &  8 & C & A & C \\ 
J182851.64+200727.2 & 12.300 &  1.28 & 0.23 & 0.05 & 1.23 & 1.67 & 0.68 &  0 &  6 & B & A & C \\ 
J182957.77+174455.2 & 12.252 &  N/A  & 0.21 & 0.16 & 1.25$^\dagger$ & 1.69$^\dagger$ & 1.15$^\dagger$ &  0 & 25 & B & A & C \\ 
J183118.99+150600.9 & 12.472 &  2.02 & 0.51 & 0.10 & 0.83 & 1.16 & 0.91 &  0 & 17 & A & A & C \\ 
\hline
\multicolumn{13}{c}{Field SW1815+0928} \\
\hline
J175913.94+132849.4 & 10.645 &  1.68 & 0.18 & 0.10 & 0.98 & 1.85 & 0.98 &  0 &  0 & C & A & A \\ 
J180202.52+065737.9 & 10.796 &  1.76 & 0.26 & 0.07 & 0.94 & 1.26 & 0.82 &  0 &  5 & A & A & C \\ 
J181222.90+100032.6 & 12.196 &  1.29 & 0.29 & 0.06 & 1.23 & 2.19 & 0.82 &  0 &  9 & C & A & C \\ 
J181858.42+103550.1 & 10.675 &  1.24 & 0.17 & 0.06 & 1.26 & 1.20 & 0.90 &  0 &  3 & A & A & C \\ 
J182127.51+094038.2 & 11.573 &  N/A  & 0.66 & 0.19 & 0.63$^\dagger$ & 0.65$^\dagger$ & 1.48$^\dagger$  &  3 & 29 & A & A & D \\ 
J182317.92+063936.2 & 12.289 &  N/A  & 0.34 & 0.09 & 0.98$^\dagger$ & 1.45$^\dagger$ & 0.85$^\dagger$ &  8 & 40 & A & A & D \\ 
J182543.36+122925.0 & 11.136 &  1.95 & 0.30 & 0.08 & 0.86 & 1.70 & 0.75 &  0 &  3 & B & A & C \\ 
J182741.05+082414.0 & 12.205 &  2.72 & 0.58 & 0.10 & 0.70 & 1.30 & 1.39 &  0 &  4 & A & A & C \\ 
\hline
\multicolumn{13}{l}{$^a$ Number of brighter objects within aperture. $^b$ No. of objects less than 5 mags fainter within aperture.} \\ 
\multicolumn{13}{l}{$^\dagger$ Based on $J-H$ colour not a $V_{SW}-K$ colour. }\\
\multicolumn{13}{l}{$^\ddagger$ Codes: R=Planet radius (A=$R_p<1.6$, B=$1.6\geq R_p> 1.75$, C=$R_p \geq1.75$),} \\
\multicolumn{13}{l}{Eta=$\eta_p$ (A=$0.5\geq \eta_p \geq 1.5$, B=$\eta_p<0.5$, C=$\eta_p > 1.5$),} \\
\multicolumn{13}{l}{B=Blending (A=OK, B=1 or 2 fainter objs in aperture, C=$>2$ fainter objs in aperture, D=brighter obj in aperture)}\\
\end{tabular}
\end{table*}

Those transit candidates that have at least 2 A's and no more than 1 B or are
close to the border between an A and a B with believable transit signals have
been designated as ``final transit candidates" and are highlighted in
\textbf{bold} in Tables~\ref{tab:newinitialcands1}--\ref{tab:filtercands1}. The
number of these final transit candidates are also shown in the last column of
Table~\ref{tab:fields}.

\begin{table*}
\centering
\caption{Final list of candidate planets. Note two objects appear twice in
different fields.}
\protect\label{tab:candplanets}
\vspace{5mm}
\begin{tabular}{lccccccccccc}
\hline
Ident. (1SWASP+)& Field & Epoch & Period & $\delta$ & Duration & $\rmn{N}_{tr}$ & $R_*$ & $R_p$ & $\eta_p$ & Sp. & Priority \\
      	      	&   & (2450000+) & (days) & (mag)  & (hrs)  &   	  & (\Rsun) & (\Rj) &	      & type &\\
\hline		
J165949.13+265346.1 & SW1717+2326 & 3128.1597 & 2.683010 & 0.0209 &  1.824 & 13 & 1.08 & 1.33 & 0.61 & G1 & 1 \\
J172549.13+502206.4 & SW1739+4723 & 3137.4856 & 4.542614 & 0.0345 &  1.536 &  8 & 0.97 & 1.54 & 0.45 & G5 & 2 \\
J173748.98+471348.7 & SW1739+4723 & 3139.1697 & 3.337786 & 0.0105 &  3.456 & 10 & 1.01 & 0.88 & 1.16 & G3 & 1 \\
J174058.24+062638.1 & SW1745+1028 & 3146.9646 & 4.804517 & 0.0168 &  4.560 &  7 & 0.82 & 0.91 & 1.50 & K0 & 2 \\
J174118.30+383656.3 & SW1741+4024 & 3138.4246 & 4.245098 & 0.0128 &  3.264 &  8 & 1.21 & 1.17 & 0.90 & F8 & 1 \\
J175401.58+322112.6 & SW1743+3126 & 3127.8124 & 1.949258 & 0.0136 &  1.992 & 20 & 0.91 & 0.91 & 0.84 & G8 & 1 \\
J175511.09+134731.5 & SW1745+1028 & 3149.3391 & 2.444503 & 0.0201 &  2.712 & 10 & 1.17 & 1.42 & 0.90 & F9 & 1 \\
J175511.09+134731.5 & SW1745+1727 & 3149.3479 & 2.443916 & 0.0242 &  2.520 & 12 & 1.17 & 1.55 & 0.83 & F9 & 1 \\
J175620.84+253625.7 & SW1744+2427 & 3124.5637 & 4.415010 & 0.0340 &  2.568 &  8 & 0.99 & 1.56 & 0.75 & G4 & 1 \\
J175856.34+421950.9 & SW1741+4024 & 3137.7993 & 3.256700 & 0.0221 &  2.088 &  8 & 1.06 & 1.34 & 0.66 & G1 & 1 \\ 
J175856.34+421950.9 & SW1744+3944 & 3127.9993 & 3.258676 & 0.0301 &  2.184 & 10 & 1.06 & 1.57 & 0.68 & G1 & 1 \\
J175914.99+213803.9 & SW1814+1727 & 3148.7092 & 4.552594 & 0.0302 &  4.224 &  4 & 0.89 & 1.32 & 1.30 & G8 & 2 \\
J180010.55+214510.2 & SW1744+2427 & 3125.6943 & 3.434074 & 0.0557 &  2.184 &  8 & 0.69 & 1.39 & 0.82 & K5 & 2 \\
\hline
\end{tabular}
\end{table*}

\begin{figure*}
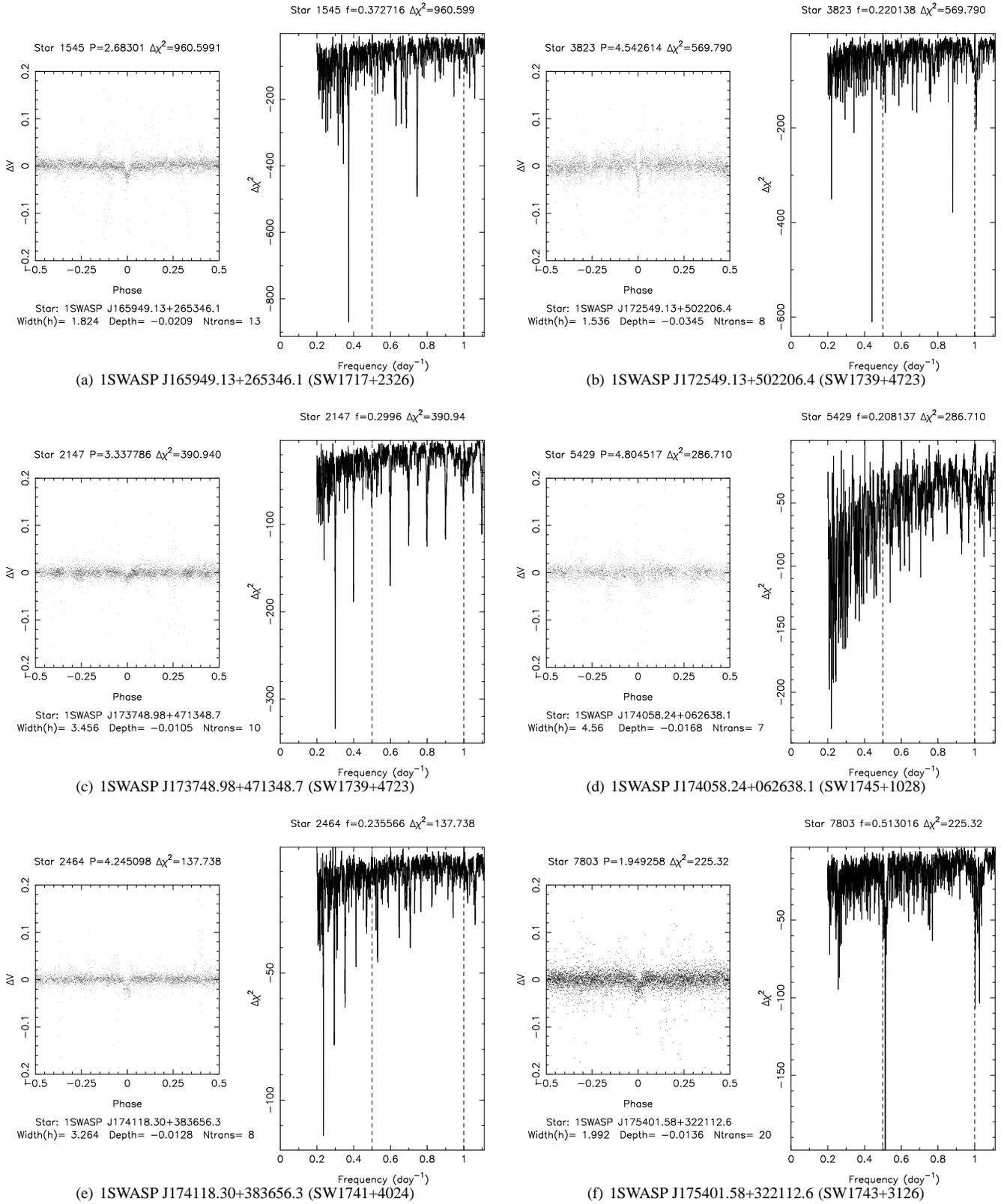

\def\subfigtopskip{0pt}    
\def\subfigbottomskip{4pt}
\def\subfigcapskip{1pt}
\centering

\begin{tabular}{cc}
\subfigure[1SWASP J165949.13+265346.1 (SW1717+2326)]{\label{fig:J165949.13_lc}
\includegraphics[angle=270,width=8.5cm]{J165949.13+265346.1_lc_big.ps}
}
&
\subfigure[1SWASP J172549.13+502206.4 (SW1739+4723)]{\label{fig:J172549.13_lc}
\includegraphics[angle=270,width=8.5cm]{J172549.13+502206.4_lc_big.ps}
} \\

\subfigure[1SWASP J173748.98+471348.7 (SW1739+4723)]{\label{fig:J173748.98_lc}
\includegraphics[angle=270,width=8.5cm]{J173748.98+471348.7_lc_big.ps}
} 
&
\subfigure[1SWASP J174058.24+062638.1 (SW1745+1028)]{\label{fig:J174058.24_lc}
\includegraphics[angle=270,width=8.5cm]{J174058.24+062638.1_lc_big.ps}
} 
\\

\subfigure[1SWASP J174118.30+383656.3 (SW1741+4024)]{\label{fig:J174118.30_lc}
\includegraphics[angle=270,width=8.5cm]{J174118.30+383656.3_lc_big.ps}
} 
&
\subfigure[1SWASP J175401.58+322112.6 (SW1743+3126)]{\label{fig:J175401.58_lc}
\includegraphics[angle=270,width=8.5cm]{J175401.58+322112.6_lc_big.ps}
} \\

\end{tabular}
\protect\caption{Light curves and periodograms for filtered transit candidates}
\label{fig:SW1717+2326trancands}
\end{figure*}

\section{Results}
\protect\label{sec:results}
\subsection{Candidate light curves}

After the blending and companion radius analysis, we are able to produce a final
list of extrasolar planetary candidates for potential follow-up. These are shown
in Table~\ref{tab:candplanets} where we list the identifier, transit parameters
(epoch, period, depth  ($\delta$) \& duration), stellar parameters (radius
($R_*$) and spectral type), derived planet parameters ($R_p$ \& $\eta_p$), along
with a suggested priority for follow-up. This priority is only based on the
analysis codes listed above and does not take into account factors such as
brightness and visibility of the target. The spectral type has been estimated
from the $V_{SW}-K$ colour and the calibration for F, G \& K dwarfs by
\cite{ammons06}.

The phase-folded light curves and periodograms for each transit candidate are
shown in Figures~\ref{fig:J165949.13_lc}--\subref{fig:J180010.55_lc}.
In the periodogram plots, the dashed vertical lines indicate the main aliased
periods of 1 and 2 days. We note the periodograms are not strict Fourier power
spectra but rather plots of \dchs\ as a function of trial period and so there is
normally little power at the traditional aliased periods of multiples of 1 day.

\subsection{Discussion of individual final candidates}

\subsubsection{1SWASP J165949.13+265346.1}

The light curve for this candidate (Figure~\ref{fig:J165949.13_lc}) is nicely
flat outside of eclipse, despite the supposedly high value of S/N$_{ellip}$ of
9.717, with a well-defined transit. There is a strong signal in the periodogram
at the period and a large number (13) of transits were detected. The measured
transit duration (1.824 hrs) is a little small for the size of the planet,
leading to a somewhat small value of $\eta_p$ of 0.61. The only objects within
the 48\arcsec\  aperture are 6 magnitudes or more fainter and make it unlikely
that blending is the cause of the low $\eta_p$ value. The candidate merits
further follow-up.

\subsubsection{1SWASP J172549.13+502206.4}

The light curve for this object (Figure~\ref{fig:J172549.13_lc}) has some
scatter and the transit shape is quite narrow and V-shaped. The inferred stellar
radius and moderately large transit depth leads to a quite large planetary
radius of 1.54\,\Rj\ with a correspondingly small value of $\eta_p$. This
combined with the relatively high value of S/N$_{ellip}=6.495$, could make the
companion a low-mass star. There is a 4.5\,mag fainter object on the edge of the
aperture 46\arcsec\ away.

\subsubsection{1SWASP J173748.98+471348.7}
This candidate has a flat light curve although the transit is quite shallow. The
estimated planetary radius is quite small at 0.88\,\Rj\ but is close to what is
expected from the duration ($\eta_p=1.16$) and probably within the errors
propagated from the colour indices. There is a 4.3\,mag fainter object towards
the edge of the aperture 43\arcsec\ away. The transit appears to be
flat-bottomed in the folded light curve and this candidate deserves
spectroscopic follow-up to determine the true nature of the object.

\subsubsection{1SWASP J174058.24+062638.1}

There is some scatter in the light curve (Figure~\ref{fig:J174058.24_lc}) and
the transit is somewhat shallow, but there is a quite strong peak in the
periodogram. The observed duration of the transit is long for an estimated
planet radius $\lesssim1\Rj$, leading to a high value of $\eta_p$ (1.50)
indicating this may be a grazing incidence stellar binary. There are two objects
4.3 and 4.9\,magnitudes fainter at a distance of 34 and 47\arcsec\ from the
candidate respectively. The combination of these factors make this object a
lower priority target.

\subsubsection{1SWASP J174118.30+383656.3}
The transit in this object is quite obvious and there is quite a strong signal
in the periodogram from the 8 detected transits. The derived spectral type of F8
leads to a quite large stellar radius but the measured transit duration is close
to what is expected ($\eta_p=0.90$) and the planetary radius is reasonable at
1.17\,\Rj. There is a very small degree of blending with a 3.5\,mag fainter
object 19\arcsec away and this could be a grazing incidence stellar binary with
similar components on twice the period. Overall this candidate would merit
further follow-up.

\subsubsection{1SWASP J175401.58+322112.6}
The light curve (Figure~\ref{fig:J175401.58_lc}) shows a fair amount of scatter
due to the relative faintness of the object ($V_{SW}\sim12.5$) but the transit
is readily visible. The period is somewhat close to the $P=2$ day alias and it
is possible the object may be a false positive, despite the large number of
detected transits. The derived planetary radius (0.91\,\Rj) and $\eta_p$ (0.84)
are both reasonable and there is no evidence of any blends. The object is worthy
of follow-up with the above caveats.

\subsubsection{1SWASP J175511.09+134731.5}

This candidate is detected in two different fields from different cameras which
overlap slightly; SW1745+1028 (DAS 1) and SW1745+1727 (DAS 2). The light curves
(Figures~\ref{fig:J175511.09_lc_a}, \ref{fig:J175511.09_lc_b}) look very similar
with a clear transit signature and the derived parameters also agree quite well.
The determined periods are very similar with small differences in the fitted
transit depth leading to small differences in the derived planetary radius
(1.42\,\Rj\  vs. 1.55\,\Rj). The derived radii are on the large side for a
planet but the calculated transit duration is close to the measured one
($\eta_p\sim1$). This candidate could be planet or possibly a brown dwarf and
warrants further follow-up.

\subsubsection{1SWASP J175620.84+253625.7}

The transit is clearly visible in the light curve
(Figure~\ref{fig:J175620.84_lc}) although it has something of a ``V-shape". The
large amplitude ($\delta-0.0340$) leads to a fairly large planet radius of
1.56\,\Rj. This may indicate that the companion is actually stellar although the
$\rmn{S/N}_{ellip}$ value is low at 2.012 and the light curve is very flat
outside of transit. Additional observations would help to secure the
exoplanetary nature of the companion.

\subsubsection{1SWASP J175856.34+421950.9}
This candidate is detected in two different fields from different cameras which
overlap slightly; SW1741+4024 (DAS 1) and SW1744+3944 (DAS 5). The light curves
(Figures~\ref{fig:J175856.34_lc_a}, \ref{fig:J175856.34_lc_b}) look very
similar and the derived transit parameters also agree quite well with only
slight differences in the derived transit depth (0.0221 in field SW1741+4024
compared to 0.0301 in SW1744+3944) leading to small differences in the derived
planetary radius (1.34\,\Rj\  vs. 1.57\,\Rj). The transit egress for the light
curve from the SW1744+3944 field is somewhat clearer and more pronounced which
probably accounts for the somewhat deeper transit amplitude fitted to these
data. The only object within the aperture is $\sim4.8$\,mag fainter and
26\arcsec\ away and this candidate would warrant spectroscopic follow-up.

\subsubsection{1SWASP J175914.99+213803.9}

There is a fair amount of scatter in this faint candidate and only 4 transits
were detected, although there is a clear peak in the periodogram. The derived
$R_p$ and $\eta_p$ are quite large for the size of the star (1.32\,\Rj\ and 1.30
respectively) and there is a 3.1\,mag fainter object $\sim19\arcsec$ away and
several other faint sources with the \SW\ aperture so blending cannot be totally
ruled as a cause of the variations.

\begin{figure*}
\addtocounter{figure}{-1}
\def\subfigtopskip{0pt}    
\def\subfigbottomskip{4pt}
\def\subfigcapskip{1pt}
\addtocounter{subfigure}{6}
\centering

\begin{tabular}{cc}
\subfigure[1SWASP J175511.09+134731.5 (SW1745+1028)]{\label{fig:J175511.09_lc_a}
\includegraphics[angle=270,width=8.5cm]{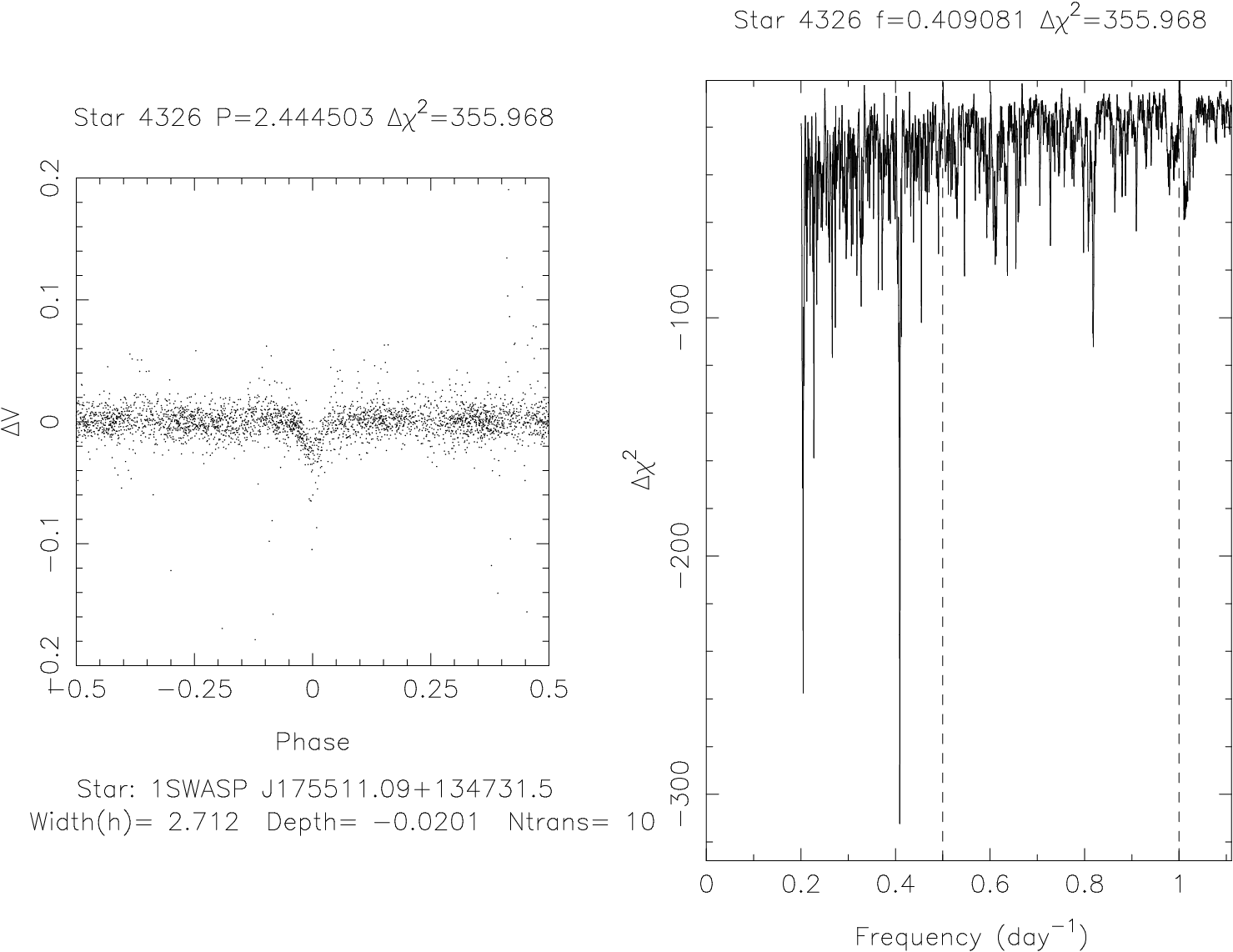}
}
&
\subfigure[1SWASP J175511.09+134731.5 (SW1745+1727)]{\label{fig:J175511.09_lc_b}
\includegraphics[angle=270,width=8.5cm]{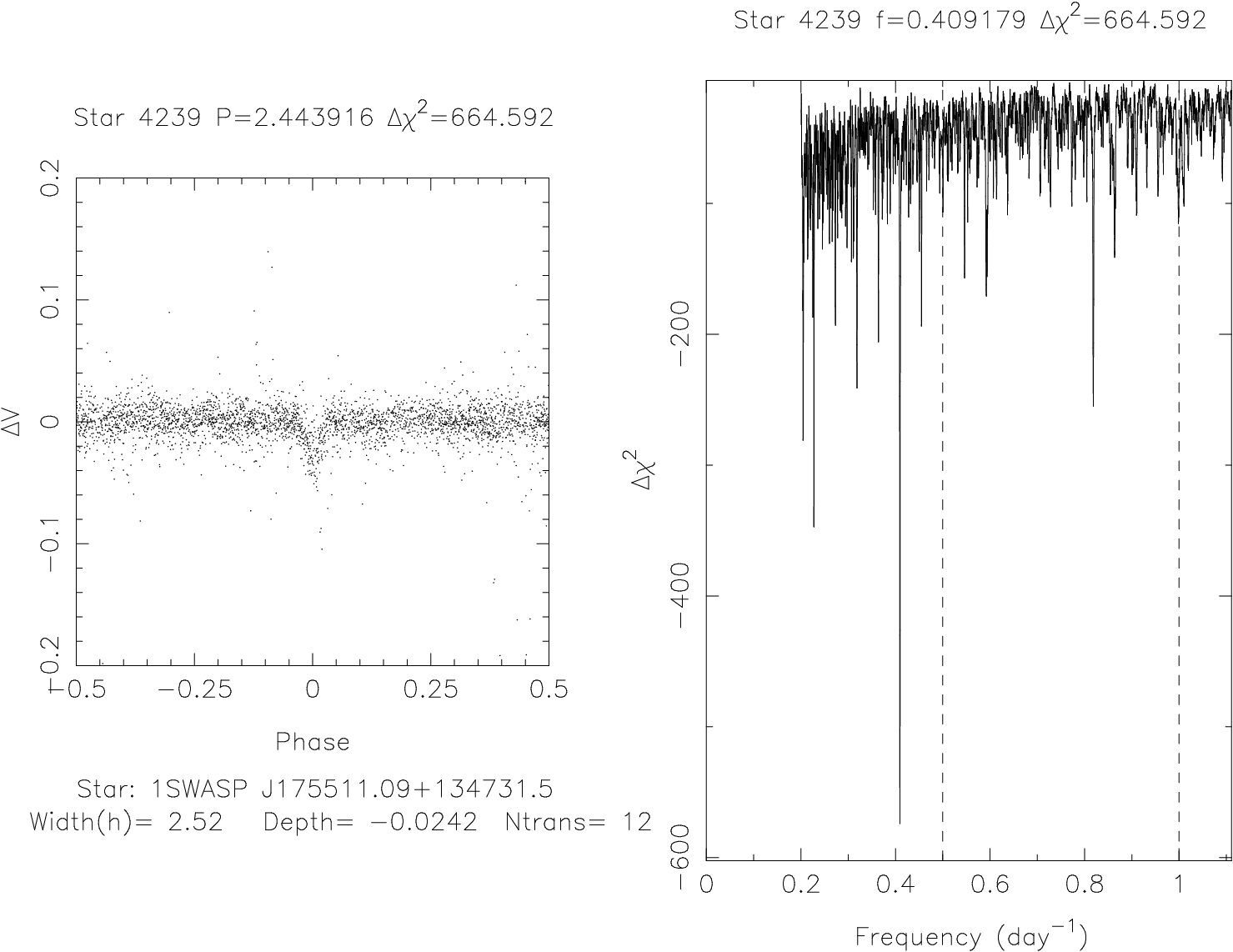}
} 
\\

\subfigure[1SWASP J175620.84+253625.7 (SW1744+2427)]{\label{fig:J175620.84_lc}
\includegraphics[angle=270,width=8.5cm]{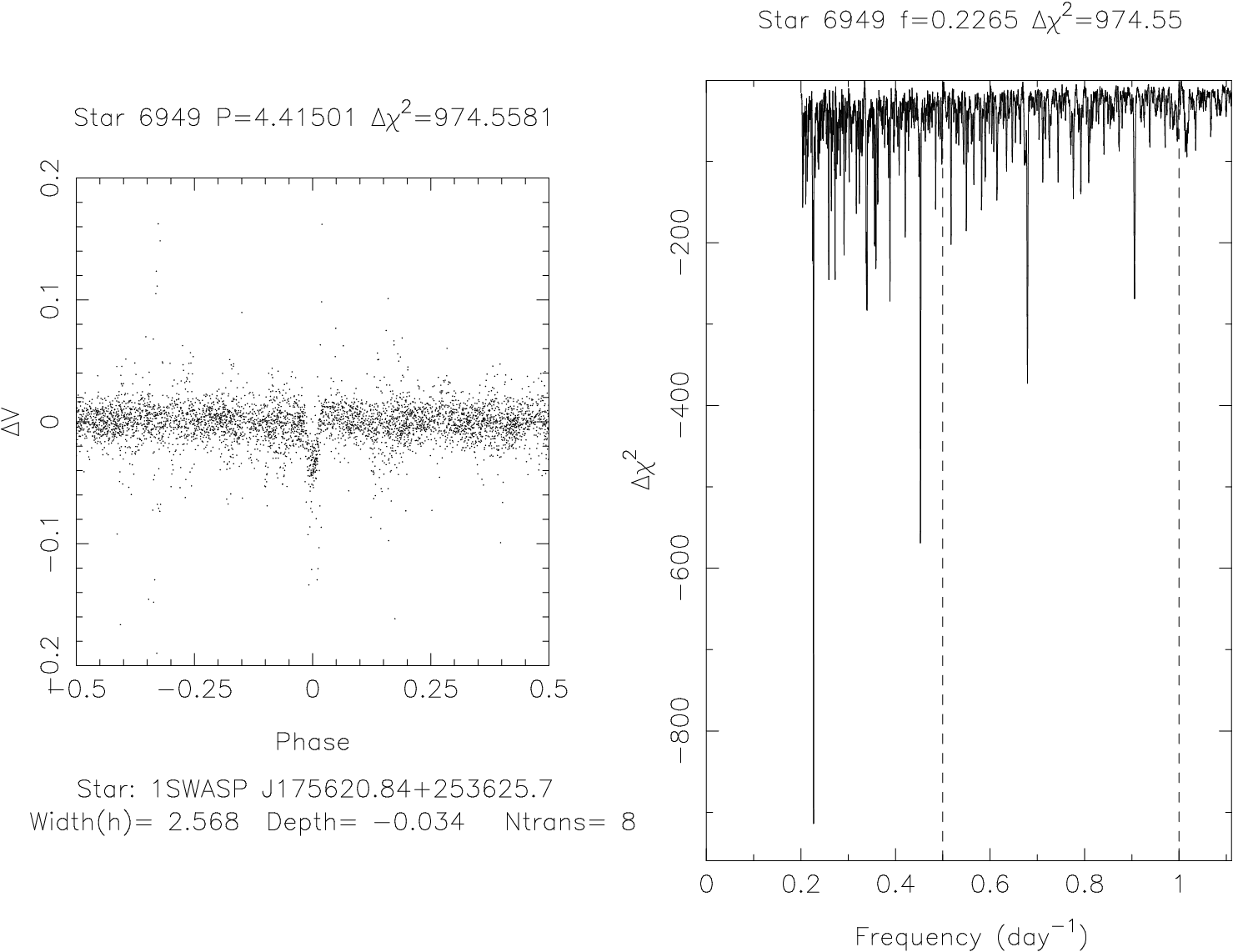}
}
& 
\subfigure[1SWASP J175856.34+421950.9 (SW1741+4024)]{\label{fig:J175856.34_lc_a}
\includegraphics[angle=270,width=8.5cm]{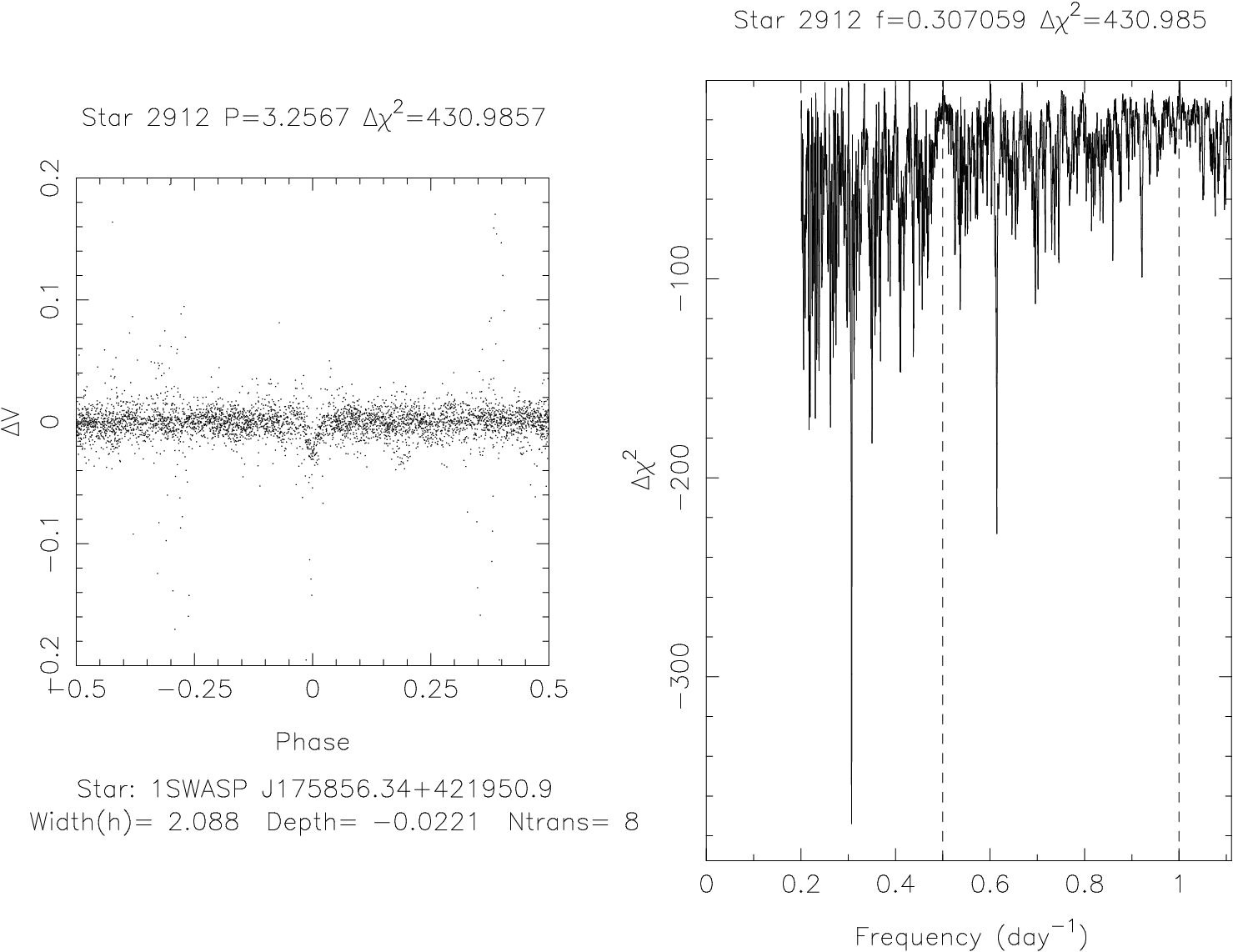}
} 
\\

\subfigure[1SWASP J175856.34+421950.9 (SW1744+3944)]{\label{fig:J175856.34_lc_b}
\includegraphics[angle=270,width=8.5cm]{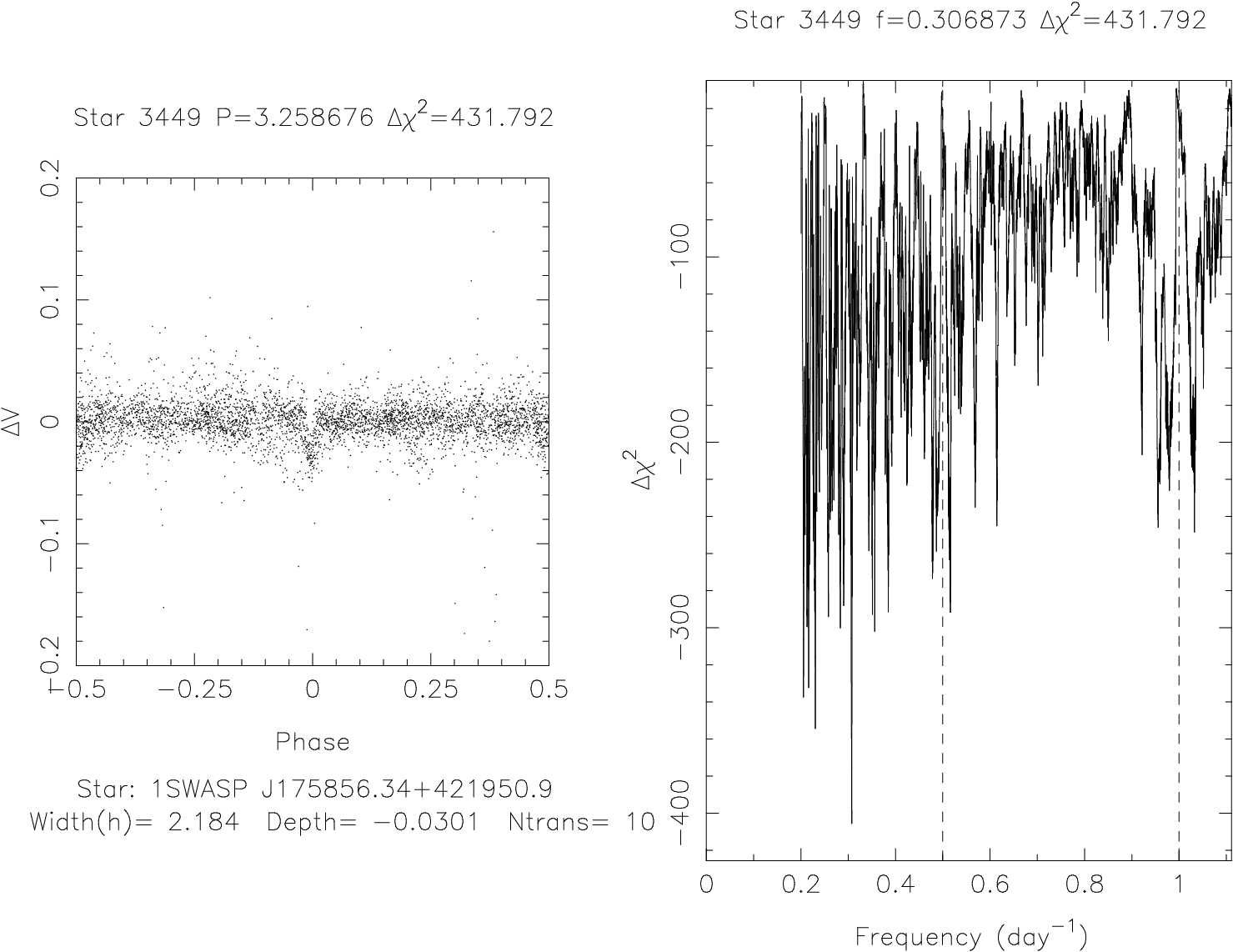}
}
&
\subfigure[1SWASP J175914.99+213803.9 (SW1814+1727)]{\label{fig:J175914.99_lc}
\includegraphics[angle=270,width=8.5cm]{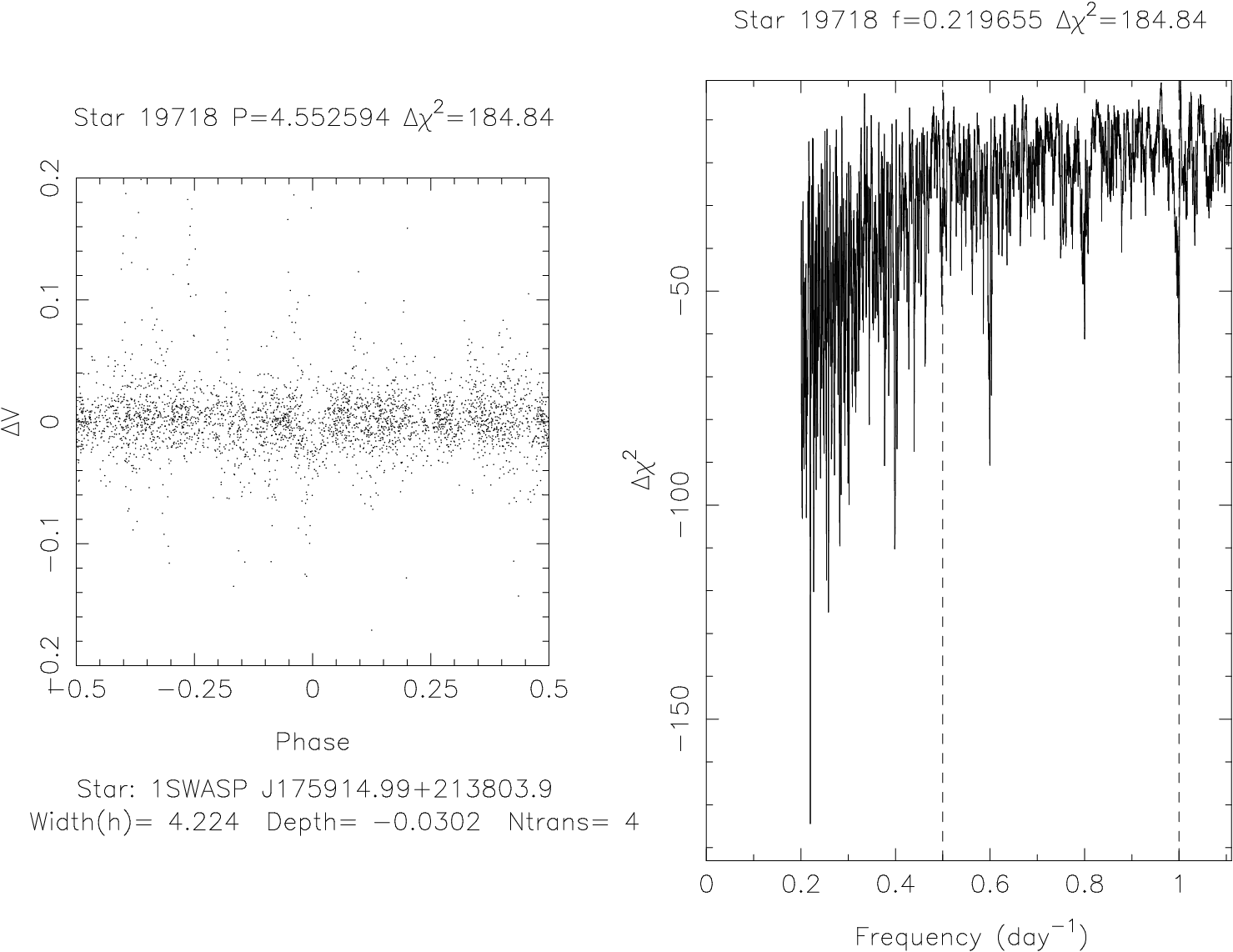}
} \\
\end{tabular}
\addtocounter{figure}{1}
\contcaption{Light curves and periodograms for filtered transit candidates}
\label{fig:SW174xtrancands}
\end{figure*}

\begin{figure*}
\addtocounter{figure}{-1}
\def\subfigtopskip{0pt}    
\def\subfigbottomskip{4pt}
\def\subfigcapskip{1pt}
\centering
\begin{tabular}{cc}

\subfigure[1SWASP J180010.55+214510.2 (SW1744+2427)]{\protect\label{fig:J180010.55_lc}
\includegraphics[angle=270,width=8.5cm]{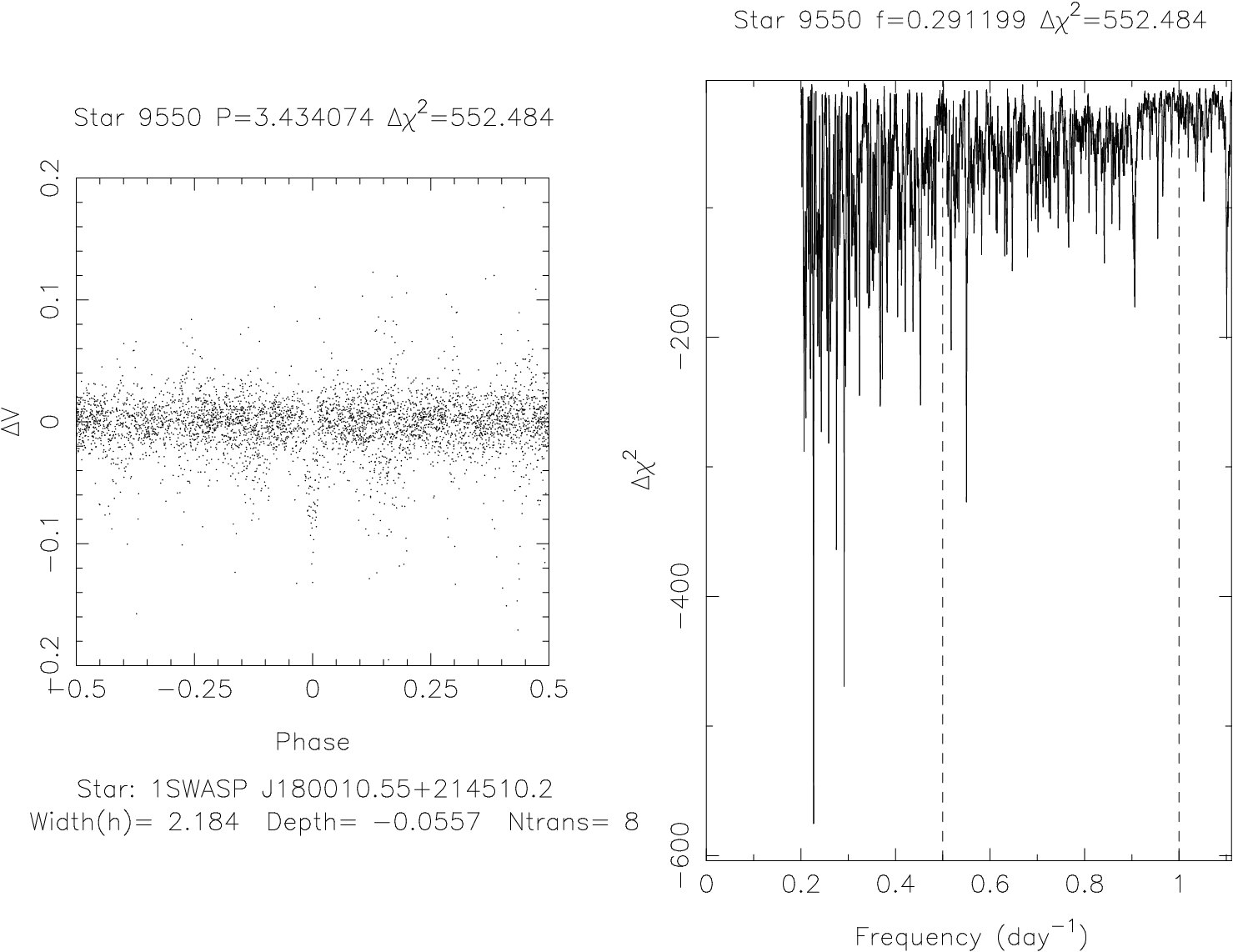}
} 
&
\subfigure[1SWASP J175143.72+205953.9 (SW1744+2427)]{\protect\label{fig:J175143.72_lc_a}
\includegraphics[angle=270,width=8.5cm]{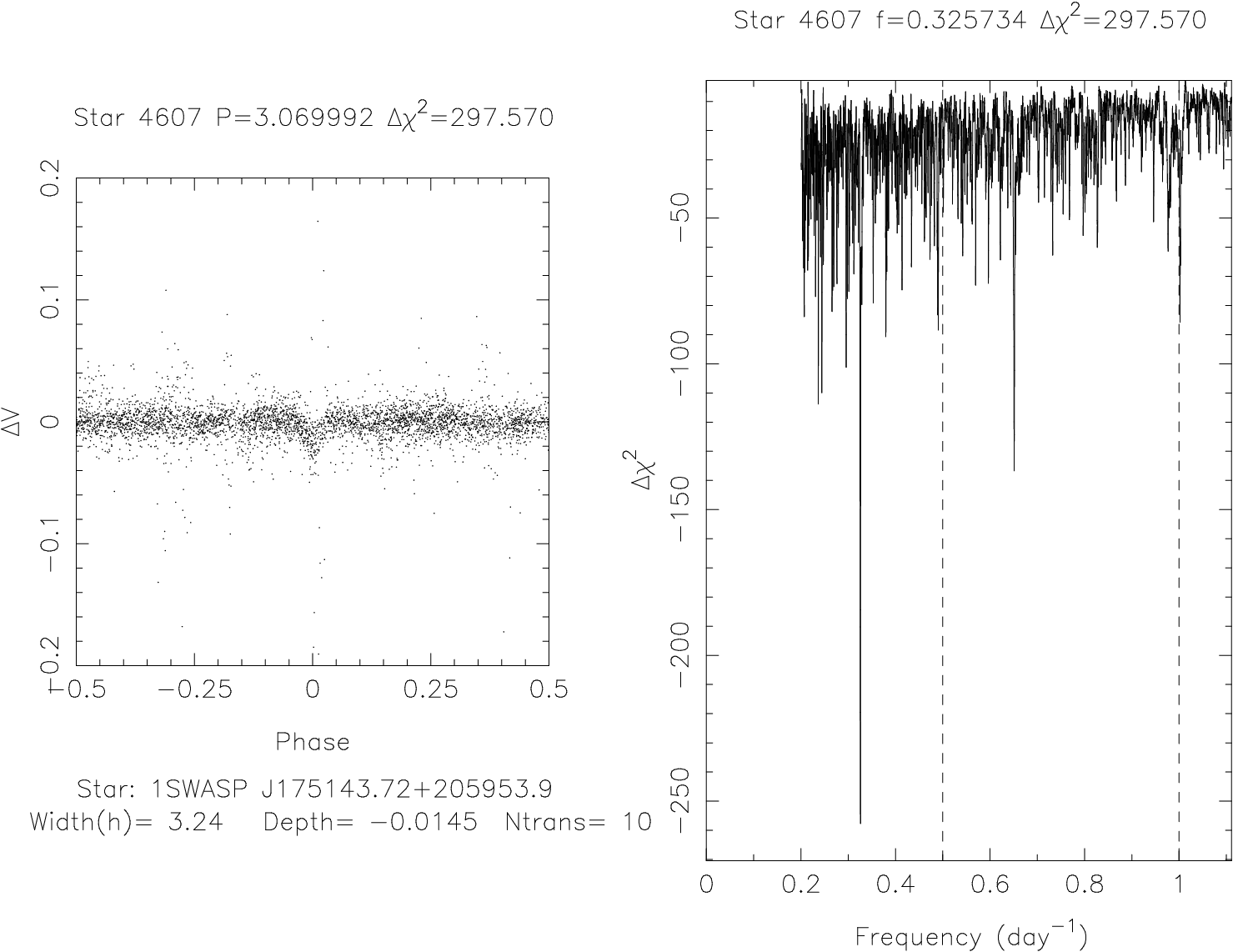}
}\\
\end{tabular}
\addtocounter{figure}{1}
\contcaption{Light curves and periodograms for filtered transit candidates}
\protect\label{fig:SW1744trancands}
\end{figure*}

\subsubsection{1SWASP J180010.55+214510.2}

This candidate has the largest depth (0.0557\,mag) of any of the candidates
although this is partly due to the late spectral type of K5 and small stellar
radius, which leads to a planetary radius of 1.39\,\Rj. The transit is somewhat
undersampled and duration is quite short but the host star has a small stellar
radius (0.69\Rsun). There are a few objects within the aperture although all are
at least 3.25\,mag fainter. Although faint and somewhat noisy (see
Figure~\ref{fig:J180010.55_lc}), this candidate orbits an interestingly late
spectral type parent star and would warrant further investigation.

\subsection{Discussion of other candidates}
\protect\label{sec:photononcands}
\subsubsection{1SWASP J175143.72+205953.9}
This object is the third that has been detected in two different overlapping
fields SW1744+2427 (DAS1) and SW1745+1727 (DAS2). The transit is readily visible
in both light curves although the data are somewhat noisy. The derived transit
parameters are very similar with depths of 0.0145 and 0.0149\,mag leading to a
rather small predicted planetary radius of 0.71 and 0.72\Rj\ when combined with
the 0.69\,\Rsun\ host star radius. 

Although the candidate has only 2 objects less than 5 mags fainter within the
48\arcsec aperture, one of these objects is $\sim3$\,mag fainter and only
6.5\arcsec away. Examining the DSS image shows an elongation of the candidate
and the 2MASS J, H \& $\rmn{K_s}$ atlas images show the close companion as an
additional source touching the candidate. With these caveats we cannot recommend
this target for follow-up although it otherwise passes all the tests. We include
the light curve from the SW1744+2427 in Figure~\ref{fig:J175143.72_lc_a} for
reference.

\subsection{Discussion of spectroscopically observed candidates}
\protect\label{sec:spectrononcands}
In this section we discuss the three objects that have been observed
spectroscopically. These objects were observed based on identification in an
earlier transit search carried our on a smaller subset of the data  but were not
selected in the more rigorous selection and filtering process presented here.
The details of the stars are included in parentheses in
Tables~\ref{tab:newinitialcands1}--\ref{tab:filtercands1} but they are not
included in the count of filtered candidates in Table~\ref{tab:fields}.

\subsubsection{1SWASP J174645.84+333411.9}

This object, which has one of the highest \dchs\  found in the transit search,
was discovered in preliminary transit searches on a small subset of the data
and was included in the list of stars observed with the CFHT and ESPaDOnS. It
was also clearly identified in this search of the full dataset, although the
high S/N$_{ellip}$ value of 18.634 means it fails the filter procedure
described in Section~\ref{sec:candfilt}. 

\begin{figure*}
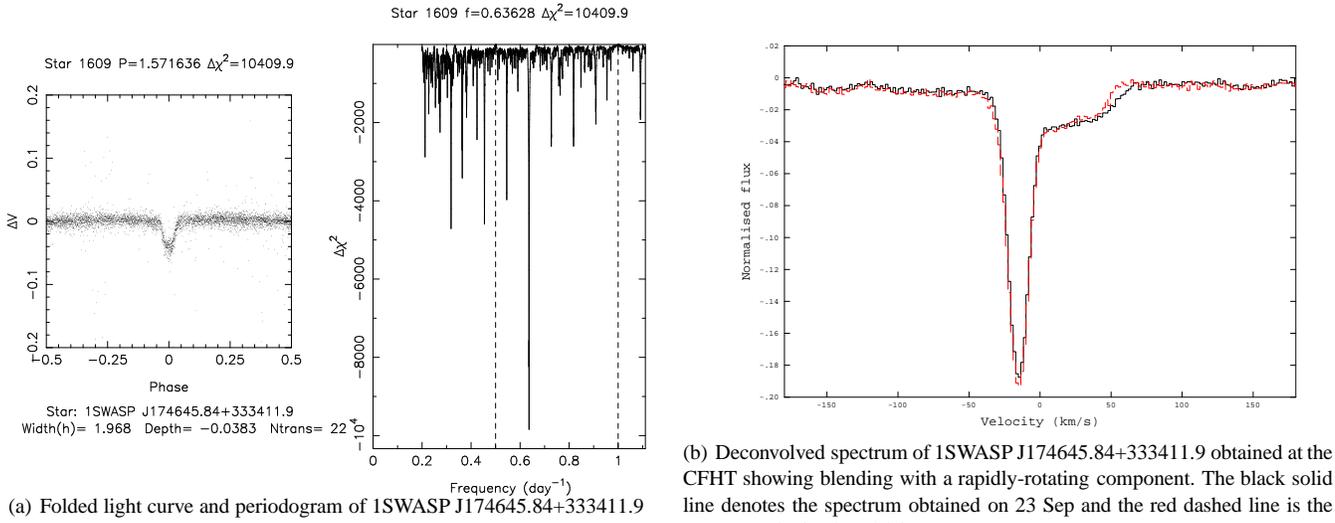

\def\subfigtopskip{0pt}    
\def\subfigbottomskip{4pt}
\def\subfigcapskip{1pt}
\setcounter{subfigure}{0}
\centering
\begin{tabular}{cc}

\subfigure[Folded light curve and periodogram of 1SWASP J174645.84+333411.9]{\label{fig:1sw1747lc}
\includegraphics[angle=270,width=8.5cm]{J174645.84+333411.9_lc_big.ps}
}
&
\subfigure[Deconvolved spectrum of 1SWASP J174645.84+333411.9 obtained at the CFHT
showing blending with a rapidly-rotating component. The black solid line denotes the spectrum obtained on 23 Sep and the red dashed line is the spectrum
obtained on 24 Sep.]{\label{fig:1sw1747spec}
\includegraphics[angle=270,width=8.5cm]{timplot.ps}
}
\end{tabular}
\protect\caption{Light curve, periodogram and deconvolved spectrum of 1SWASP
J174645.84+333411.9}
\end{figure*}

The light curve and periodogram are shown in Figure~\ref{fig:1sw1747lc} and
shown the clear transit signature and strong periodogram peak which lead to the
initial selection. The deconvolved profile is shown in
Figure~\ref{fig:1sw1747spec} and clearly shows blending with a rapidly-rotating
component. The derived colour and radius of the primary star and inferred
companion radius from Table~\ref{tab:filtercands1} of 1.94\ \Rj($\simeq 0.2$\
\Rsun) indicate that this object is likely to be a short period late F+M
stellar binary.

This object (as BD+33$^\circ$ 2954) was also detected by the HAT project
(\citealt{bakos04hat}) who also ruled it out as a transit due to the depth of
the transits and low-amplitude sinusoidal variation. They also concluded the
companion was likely to be a M dwarf.

\subsubsection{1SWASP J172826.46+471208.4}

\begin{figure*}
\def\subfigtopskip{0pt}    
\def\subfigbottomskip{4pt}
\def\subfigcapskip{1pt}
\setcounter{subfigure}{0}
\centering
\begin{tabular}{cc}

\subfigure[Folded light curve and periodogram of 1SWASP
J172826.46+471208.4]{\label{fig:1sw1728lc}
\includegraphics[angle=270,width=8.5cm]{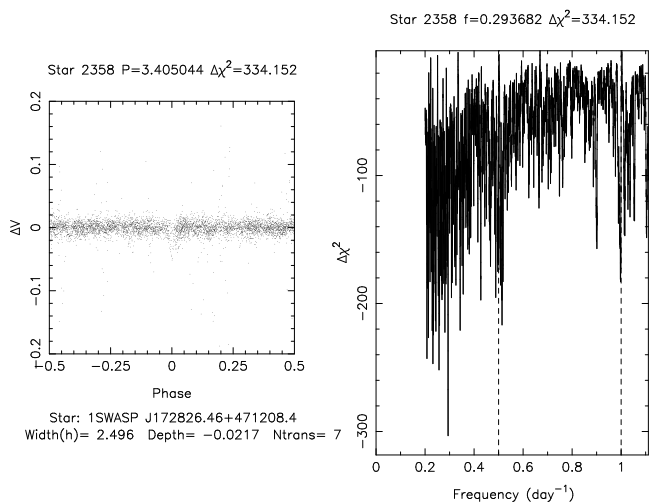}
}
&
\subfigure[Deconvolved spectrum of 1SWASP J172826.46+471208.4 obtained at the CFHT
clearly showing a double-lined spectrum.]{\label{fig:1sw1728spec}
\includegraphics[angle=270,width=8.5cm]{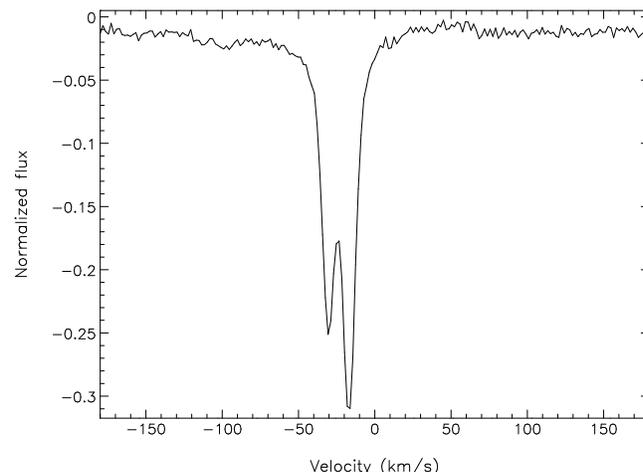}
}
\end{tabular}
\caption{Light curve, periodogram and deconvolved spectrum of 1SWASP J172826.46+471208.4
}
\end{figure*}

This object was identified as a possible candidate during visual inspection in
the full dataset but failed the selection cuts described in
Section~\ref{sec:candfilt} due to only having  \dchs/\dchs$_-=1.13$, less than
the 2.0 required. The deconvolved profile is shown in
Figure~\ref{fig:1sw1728spec} and clearly shows a double-lined signature
indicating that it is a stellar binary.

The derived stellar and planetary parameters are shown in
Table~\ref{tab:filtercands1} for comparison purposes and indicate the value of
the high resolution spectrum. The planetary radius (1.03\,\Rj) and $\eta_p$
value of 0.91, combined with the absence of any brighter or fainter companions
would otherwise lead to it have been considered a good candidate.

\subsubsection{1SWASP J173403.61+280145.1}

\begin{figure*}
\def\subfigtopskip{0pt}    
\def\subfigbottomskip{4pt}
\def\subfigcapskip{1pt}
\setcounter{subfigure}{0}
\centering
\begin{tabular}{cc}

\subfigure[Folded light curve and periodogram of 1SWASP
J173403.61+280145.1]{\label{fig:1sw1734lc}
\includegraphics[angle=270,width=8.5cm]{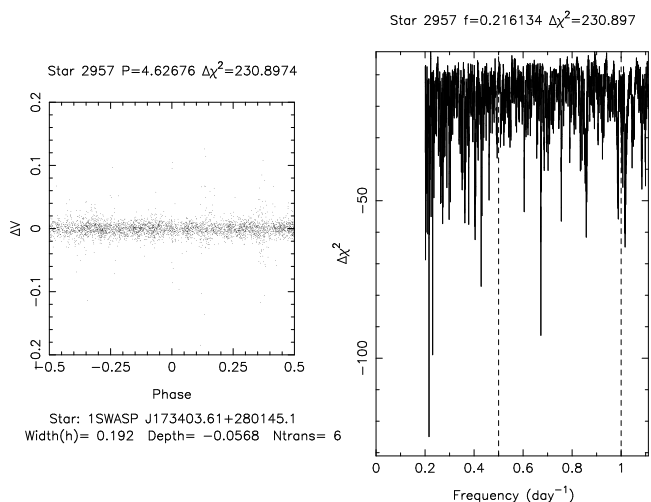}
}
&
\subfigure[Deconvolved spectrum of 1SWASP J173403.61+280145.1 obtained at the
CFHT.]{\label{fig:1sw1734spec}
\includegraphics[angle=270,width=8.5cm]{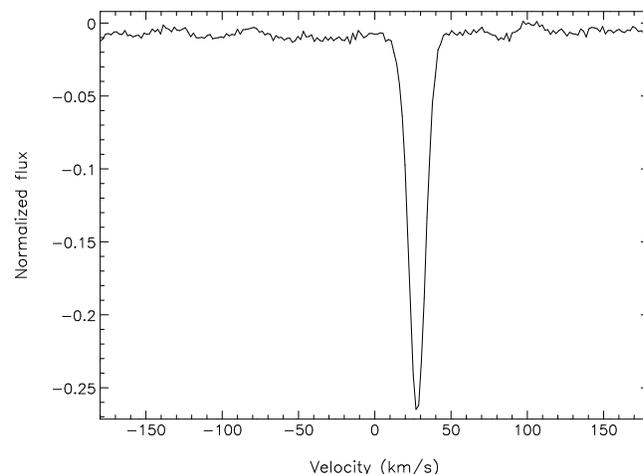}
}
\end{tabular}
\caption{Light curve, periodogram and deconvolved spectrum of 1SWASP J173403.61+280145.1}
\end{figure*}

This object was found in the preliminary transit searches but was not
identified as a potential candidate in the full dataset as there was no
convincing evidence of a transit (see Figure~\ref{fig:1sw1734lc}). The reported best period this time
($P=4.62676\ \rmn{days}$) is considerably different from the previous period
($P=3.72987\ \rmn{days}$) based on only 2 transits, indicating the previous
period was probably spurious. There is also no evidence of a periodic signal at
the previously detected period with $\dchs>80$ in any of the top five peaks.  In
addition, the fitted transit duration is unphysically short (0.192 hours) and
no evidence of a transit dip is seen in the folded light curves at any of the
five alternatives period reported by \textsc{hunstman}. The deconvolved profiles
are shown in Figure~\ref{fig:1sw1734spec} and shows a narrow-lined spectrum but
with only a single epoch of observation, radial velocity variation cannot be
ruled out. It is probably a non-transiting single star.

\section{Discussion}
\protect\label{sec:discuss}

\subsection{Expected number of transit candidates}
From our sample of $\sim920,000$ stars, approximately 40\% have $V\leq13$
enabling \SW\  to detect a transit, leaving us with $\sim370,000$ stars. Of
these, only a fraction ($\sim46$\% - \citealt{smith06}) will have late enough
spectral types (F--K) and therefore small enough stellar radii to allow transits
to be detected. It is well known that the planet fraction increases with
increasing metallicity for solar neighbourhood stars (e.g. \citealt{fischer05})
but the metallicities are unknown for our field stars. Assuming a planet
fraction of 1\% and a 10\% chance of seeing transits from geometric arguments
(\citealt{horne03}) we obtain 170 transit candidates assuming 100\%
observational coverage. The mean number of nights of observation from
Table~\ref{tab:fields} is $\sim114$, and from examining
Figure~\ref{fig:SW1745+1727trancov} we can see that we recover $>80$\% of
transit candidates with $P<5$\,days if we require 4 transits to be detected.
This drops sharply to $\sim20\%$ for $4>P>5$\,days if we require 6 transits
which is likely to be a minimum number to guarantee a good detection given the
presence of correlated noise in the light curves (\citealt{smith06},
\citealt{pont06rednoise}). The combination of this factor with others related to
the finite observing window and weather effects will reduce the estimate of 170
transit candidates down to $\sim20$--30, in reasonable agreement with the
initial number (\ntr) found given the large uncertainties inherent in the above
analysis.

Comparing the transit recovery fraction from a field with 130 nights of
observations (e.g. Field SW1743+3126; see Table~\ref{tab:fields} and
Figure~\ref{fig:SW1743+3126trancov}) and $\sim90$ nights (e.g Field SW1745+1028;
see Table~\ref{tab:fields} and Figure~\ref{fig:SW1745+1028trancov}) shows that
the fraction recovered drops from $\sim80\%$ (for $4>P>5$\,days) to $<10\%$.
This indicates that it will be necessary to re-observe the same fields covered
in the 2004 season in order to increase the observing baseline and the number of
detected transits and reduce the correlated systematic noise. The impact of this
correlated noise is that fields need to be observed for much longer and
substantially more transits are needed for a secure detection than has
previously been estimated from simulations which assumed white noise. This will
lead to a much lower planet candidate yield for a transit detection experiment
than has been thought previously.

\subsection{Blending}

An additional question is what fraction of our 10 extrasolar planet candidates
will turn out to be genuine extrasolar planet candidates after additional
follow-up. Many authors have discussed the effects of blending and
contamination by eclipsing binaries in both ``wide \& shallow'' and ``narrow \&
deep'' transit searches and recently \cite{brown03} have estimated the
contamination rate to be as high as 9 out of 10 candidates. 

Through our use of filtering on the signal to red noise ratio, the transit to
anti-transit fit ratio and the amount of ellipsoidal variation, combined with
the higher resolution 2MASS atlas images to assess blending, we hope to have
eliminated the majority of blends caused by grazing incidence and low-mass
stellar binaries. The remaining category of blends identified by
\cite{brown03}, namely eclipsing binaries diluted by the light of a foreground
or background star, are more difficult to eliminate as has been shown by several
authors for different transit search projects (\citealt{torres04},
\citealt{mandushev05}, \citealt{odonovan06}). 

Our strategy of obtaining one or two high resolution spectroscopic snapshots to
rule out blends, begun in a limited fashion as described in
Section~\ref{sec:obsspec}, will be able to eliminate this category of false
positives more efficiently than through the use of multicolour photometry.
Several of the potential candidates could be eliminated with a single
$\sim10$\,min exposure when the broadened or double-lined nature was discovered.
This compares very favourably with the many hours of high precision multicolour
photometry needed to eliminate transit candidates on the basis of unequal
eclipse depths in different passbands.

\section{Conclusions}
\protect\label{sec:conclusion}

We have conducted a transit search on 13 fields in the RA range 17--18\,hrs and
extracted $\sim186,000$ stars for transit searching. From these stars we find
9847 initial transit candidates, with 199 of these passing visual inspection.
Following filtering this number was reduced to \ntr\ and with analysis of the
blending and the region around each star and the estimated planetary radii this
number is reduced to 11 extrasolar planet candidates, with 2 candidates being
detected twice in separate fields.

Initial spectroscopic follow-up on 2 candidates (1SWASP J174645.84+333411.9 and
1SWASP J172826.46+471208.4) which failed the newer, more stringent filtering in
Section~\ref{sec:candfilt} confirmed the effectiveness of this filtering as
these candidates were clearly identified as stellar binaries from the spectra. 

These results have been obtained for $\sim\!1/6^\rmn{th}$ of the total number of
stars observed by the 5 cameras of \SW-North over a period of 5 months. For the
2006 observing season we will be operating with a total of 16 cameras split over
two observing sites. In addition, we expect to be operating the instruments for
a greater fraction of the year than in the initial 2004 season, leading to a
much greater potential planet catch.

With the large increase in transiting planets expected from operating 3 times as
many cameras, we can look forward to a situation where meaningful statistical
comparisons and discrimination between potential planetary models and theories
can be made.

\section*{Acknowledgements}
The WASP consortium consists of representatives from the Queen's University
Belfast, University of Cambridge (Wide Field Astronomy Unit), Instituto de
Astrofisica de Canarias, Isaac Newton Group of Telescopes (La Palma), 
University of Keele, University of Leicester, Open University, and the
University of St Andrews. The \SW-North instrument was constructed and operated
with funds made available from the Consortium Universities and the Particle
Physics and Astronomy Research Council. \SW-North is located in the Spanish
Roque de Los Muchachos Observatory on La Palma, Canary Islands which is
operated by the Instituto de Astrof\'isica de Canarias (IAC). The data
reduction and analysis described in this made made extensive use of the
Starlink Software Collection, without which this project would not have been
possible. This research also made use of the \textsc{simbad} database and
\textsc{vizier} catalogue service, operated at CDS, Strasbourg, France. In
addition we made use of data products from the Two Micron All Sky Survey, which
is a joint project of the University of Massachusetts and the Infrared
Processing and Analysis Center/California Institute of Technology, funded by
the National Aeronautics and Space Administration and the National Science
Foundation.

\bibliographystyle{mn2e}
\bibliography{iau_journals,master,ownrefs}

\bsp
\label{lastpage}

\end{document}